\documentclass[preprintnumbers,floatfix,aps]{revtex4}
\usepackage{amsmath}
\usepackage{amsfonts}
\usepackage{amssymb}
\usepackage{graphicx}
\usepackage{epsfig}

\usepackage{dcolumn}
\usepackage{bm}
\usepackage{multirow}
\usepackage{float}
\usepackage{hyperref}
\usepackage{epstopdf}

\usepackage[utf8]{inputenc}

\newcommand{\Punkte}{0}

{\noindent {\bf Exercise {#1}.} #2 \vspace{0.2cm} \\  
\renewcommand{\Punkte}{#3}}%
{\mbox{\hspace{3ex}} \hfill {\bf \Punkte~\mbox{Points}}\bigskip }

\newenvironment{Exercise*}[2]%
{\noindent {\bf Exercise {#1}*.} #2 \vspace{0.2cm} \\ 
renewcommand{\Punkte}{#2}}%
{\mbox{\hspace{2ex}} \hfill {\bf \Punkte~\mbox{Points}}\bigskip }

{\noindent {\bf Exercise {#1}.}\vspace{0.2cm} #2}%
{}

\newcounter{enum1}

\newcounter{enuma}

\begin{document}

\title{{\bf Bright Discrete Solitons in Spatially Modulated 
DNLS Systems}}
\date{\today}

\author{P. G. Kevrekidis \thanks{%
Email: kevrekid@math.umass.edu}}
\affiliation{Department of Mathematics and Statistics, University of Massachusetts,
Amherst, MA 01003-4515, USA}

\affiliation{Center for Nonlinear Studies and Theoretical Division, Los Alamos
National Laboratory, Los Alamos, NM 87544}

\author{R. L. Horne}
\affiliation{Department of Mathematics, Morehouse College, Atlanta, GA 30314, USA}

\author{N. Whitaker}
\affiliation{Department of Mathematics and Statistics, University of Massachusetts,
Amherst, MA 01003-4515, USA}

\author{Q. E. Hoq}
\affiliation{Department of Mathematics, Western New England University, Springfield, Massachusetts 01119, USA}

\author{D.\ Kip}
\affiliation{Faculty of Electrical Engineering, Helmut Schmidt University, 22043 Hamburg, Germany}

\begin{abstract}
In the present work, we revisit the highly active research 
area of inhomogeneously nonlinear defocusing media and consider
the existence, spectral stability and nonlinear dynamics of
bright solitary waves in them. We use the anti-continuum limit
of vanishing coupling as the starting point of our analysis, 
enabling in this way a systematic characterization of the branches
of solutions. Our stability findings and bifurcation characteristics
reveal the enhanced robustness and wider existence intervals
of solutions with a broader support, culminating in the 
``extended'' solution in which all sites are excited. 
Our eigenvalue predictions are corroborated by numerical
linear stability analysis. Finally, the dynamics also reveal
a tendency of the solution profiles to broaden, in line with the
above findings. These results pave the way for further explorations
of such states in discrete systems, including in higher dimensional
settings.
\end{abstract}

\maketitle



\section{Introduction \& Background}

Intrinsic localized modes have received considerable attention
during the past two and a half decades since their theoretical
inception; see e.g.~\cite{sievtak,page}. During this time frame,
a wide range of experimental contexts has been shown to support
their existence and robust role in the systems' dynamics.
These range from arrays of nonlinear-optical waveguides \cite{moti} to 
Bose-Einstein condensates (BECs) in periodic potentials \cite{ober},
and from
micromechanical cantilever arrays~\cite{sievers} to
Josephson-junction ladders \cite{alex}. 
Additional areas of application include (but are not limited to)
granular crystals of beads interacting through Hertzian contacts~\cite{theo10},
layered antiferromagnetic crystals~\cite{lars3}, as well as
halide-bridged transition metal complexes~\cite{swanson},
 and dynamical models of the DNA double strand \cite{Peybi}.
This broad span of activities has been summarized in~\cite{review},
although admittedly applications keep being added every year;
see e.g.\ as recent examples~\cite{darkbreath,niemi}.

A model that has contributed considerably to advancing our 
understanding of such lattice nonlinear dynamical systems
and their coherent structures is the 
so-called
discrete nonlinear Schr{\"o}dinger (DNLS) equation~\cite{book}. 
Its apparent simplicity as a prototypical system incorporating
the interplay of nonlinearity and a discrete form of dispersion,
but also its  relevance as a suitable approximation
of optical waveguide systems~\cite{dnc,moti} and atomic systems in optical 
lattices~\cite{ober} have, undoubtedly, contributed to the
popularity of the model. Another key feature is its 
ability to capture numerous
linear and nonlinear features observed in experiments
such as discrete diffraction~\cite{yaron} and its management~\cite{yaron1},
discrete solitons~\cite{yaron,yaron2} and vortices~\cite{neshev,fleischer}, 
Talbot revivals~\cite{christo2}, and $\mathcal{PT}$-symmetry
and its breaking~\cite{kip}, among many others. 

On the other hand, a theme that has met with growing interest
in nonlinear Schr{\"o}dinger settings recently is that of 
spatially modulated nonlinearities; see e.g. for a review
geared specifically towards periodic modulations the work of~\cite{borisreview}.
Within that theme, a focal point has been the study of the potential
bright solitons in the context of defocusing nonlinearity, as supported
by spatial modulations. This possibility introduced for 
solitons and vortices in~\cite{malom1,malom2,malom3}
was also extended to the setting of spatially inhomogeneous
nonlinear losses in~\cite{malom4}, quintic nonlinearities
in~\cite{malom5}, domain walls in~\cite{malom6}, Fermi and
Bose gases in~\cite{malom7}, dipolar Bose-Einstein condensates
in~\cite{malom8}, nonlocal media~\cite{malom9}, discrete
systems in~\cite{malom10} and even in 3d media in~\cite{malom11}.

In the present work, we combine the two above themes. In particular,
we explore the defocusing DNLS equation in the presence of a growing
nonlinearity. Contrary to the case of~\cite{malom10}, here the
nonlinearity does not present an exponential growth, but rather a 
power law in our concrete implementation. Nevertheless, this is not
the focus of our contribution. Instead, we keep the analysis as general
as possible, considering an arbitrary profile in the nonlinearity
(given by $g(n)$) whenever possible. Our aim is to start from the
well-established anti-continuum limit of~\cite{mackay} and
following the existence and stability considerations of~\cite{pkf}
to provide a systematic view of the possible excitations in the
discrete system in the form of bright solitary waves. We examine
states with 1, 2 and 3 sites, as well as the ``extended'' state
where all the sites are excited. We reveal the stability
of the different states, and also explore how approaching the
continuum limit, more and more extended states are favored 
(while ``narrower'' states disappear in suitable bifurcations),
leaving as most suitable state in the limit the extended state
mentioned above. We also provide a comparison of the stability
properties with the corresponding homogeneous limit (where the nonlinearity
strength is equal for all sites), illustrating
that the inhomogeneous nonlinear profile effectively promotes
the instability of the few-site localized states considered.

Our presentation will be structured as follows. In section 2,
we provide the theoretical background, analyzing the existence
and stability properties of different states. In section 3,
we provide numerical existence/linear stability results that
corroborate the analysis, as well as direct numerical simulations
following the dynamics of unstable states. Finally, in section 
4, we summarize our findings and present a number of directions
for future consideration.

\section{Theoretical Analysis}%

\noindent 
The model that we will consider will be of the form:
\begin{eqnarray}
i \dot{u}_n=- \epsilon \left( u_{n-1} + u_{n+1} - 2 u_n \right)
+ g(n) |u_n|^2 u_n.
\label{ddnls1}
\end{eqnarray}
Here, we have in mind a defocusing nonlinearity, as is 
e.g. the case in LiNbO$_3$~\cite{kip1,kip2,kip3}, used
previously to demonstrate a number of features including
dark and dark-bright solitary waves.
However, the waveguides in the present setting are effectively 
``tailored'' to have distinct Kerr response, forming
the profile associated with $g(n)>0$.

We will seek standing waves in the customary form
$u_n=e^{-i \Lambda t} v_n$, ($t$ here plays the role
of the --spatial-- evolution variable and $\Lambda$ is
assumed to be positive), hence $v_n$
will satisfy:
\begin{eqnarray}
\Lambda v_n = - \epsilon \Delta_2 v_n + g(n) |v_n|^2 v_n.
\label{ddnls2}
\end{eqnarray}
Here $\Delta_2$ plays the role of the discrete Laplacian.
At the anti-continuum (AC) 
limit~\cite{mackay} of $\epsilon \rightarrow
0$, the only solutions are $v_n=0$ and $v_n= \sqrt{\Lambda/g(n)} e^{i \theta_n}$.
Enforcing the latter for every site and provided that $g(n)$
grows indefinitely leads to a decaying pulse (the extended solution
considered herein). While we will briefly touch upon this waveform,
our emphasis will be (similarly to~\cite{pkf}) on few-site excitations.

Using Eq.~(\ref{ddnls2}) multiplied by the conjugate $v_n^{\star}$
and subtracting from it the conjugate of  Eq.~(\ref{ddnls2}) multiplied
by $v_n$, we obtain a solvability condition which is the same
as in the standard DNLS case, namely:
\begin{eqnarray}
0= \dots = 
v_{n-1} v_n^{\star}- v_{n-1}^{\star} v_n = v_n v_{n+1}^{\star} - v_{n}^{\star} 
v_{n+1} = \dots = 0,
\label{ddnls3}
\end{eqnarray}
due to our implicit assumption of $|v_n| \rightarrow 0$, as $n \rightarrow
\infty$.
Using the AC limit solution of  $v_n= \sqrt{\Lambda/g(n)} e^{i \theta_n}$, this
yields that for adjacent excited sites the condition
$\sin(\theta_{n+1}-\theta_n)=0$ should hold allowing relative phases
{\it only} of $0$ or $\pi$ for such sites. 

We now explore the corresponding linearization problem, using the
ansatz
\begin{eqnarray}
u_n=e^{-i \Lambda t} \left[ v_n + \delta 
\left(p_n e^{\lambda t} + q_n^{\star} e^{\lambda^{\star} t} \right) \right]
\label{ddnls4}
\end{eqnarray}
(where $^{\star}$ denotes complex conjugate) 
and deriving the equations to O$(\delta)$ for $(a_n,b_n)$
such that $p_{n}=a_{n}+i b_{n}$ and $q_{n}=a_{n} - i b_{n}$, given
the complex nature of the perturbations to the solution
$v_n$~\cite{pkf}. Notice that
hereafter, we will restrict ourselves (without loss of
generality for the one-dimensional setting) to real solutions, assuming
$\theta_n=0$ or $\pi$. 
Here we are effectively using the gauge invariance of the DNLS to
fix one of the excited sites' phase to $0$ (or $\pi$) and
the solvability condition above to obtain that all other excited
site phases should then also be $0$ or $\pi$~\cite{book}.
Then, the resulting eigenvalue problem 
reads:
\begin{eqnarray}
\lambda 
\left( \begin{array}{c}
a_n \\
b_n \\
    \end{array} \right) =
\left( \begin{array}{cc}
0 & {\cal L}_- \\
- {\cal L}_+  & 0 \\
    \end{array} \right) 
\left( \begin{array}{c}
a_n \\
b_n \\
    \end{array} \right).
\label{ddnls5}
\end{eqnarray}
Here ${\cal L}_- b_n = -\epsilon \Delta_2 b_n - \Lambda b_n + g(n)
v_n^2 b_n$, while 
${\cal L}_+ a_n=-\epsilon \Delta_2 a_n - \Lambda a_n + 3 g(n)
v_n^2 a_n$. Combining the two linearization equations, we obtain
\begin{eqnarray}
\lambda^2 b_n =- {\cal L}_+ {\cal L}_- b_n
\Rightarrow \lambda^2 {\cal L}_+^{-1} b_n =-{\cal L}_- b_n.
\label{ddnls6}
\end{eqnarray}
Notice that in the vicinity of $\epsilon \rightarrow 0$, 
${\cal L}_+$ becomes a multiplicative operator with positive
entries, hence is invertible. Now, forming the inner product
with $b_n$, we obtain
\begin{eqnarray}
\lambda^2 =-\frac{ \langle b_n, {\cal L}_- b_n \rangle}{
\langle b_n, {\cal L}_+^{-1} b_n \rangle}.
\label{ddnls7}
\end{eqnarray}
But again, near the AC limit $v_n^2 \rightarrow \Lambda/g(n)$,
leading to ${\cal L}_+ \rightarrow 2 \Lambda$ and hence
${\cal L}_+^{-1} \rightarrow (2 \Lambda)^{-1}$. Therefore, the
stability will critically hinge on the eigenvalues of
${\cal L}_-$. 

${\cal L}_-$ can be directly seen
by the considerations above to vanish at the AC limit
for the {\it excited sites}. For the {\it non-excited sites},
${\cal L}_+  = {\cal L}_- = - \Lambda$, yielding $\lambda= \pm \Lambda i$.
Hence, at the AC limit, except for the $N$ excited sites corresponding
to $0$ eigenvalues, all other eigenvalues will be degenerate
at  $\pm \Lambda i$ and as $\epsilon$ becomes nonzero will form
the continuous spectrum  $[\Lambda - 4 \epsilon, \Lambda]$;
hereafter and without loss of generality we will
set $\Lambda=1$. However, the 
key for stability considerations regards the $N-1$ eigenvalue
pairs bifurcating from the origin in the case of $N$ excited sites
(one pair will stay at  $\lambda=0$ due to the gauge invariance of
the model). To determine these eigenvalues, it is critical
to evaluate the $N \times N$ reduction of the
operator ${\cal L}_-$ so as to obtain the eigenvalues from
Eq.~(\ref{ddnls7}). To do so, we follow a similar approach
as in~\cite{pkf} expanding $v_n= v_n^{(0)} + \epsilon v_n^{(1)} + 
\dots$ and computing the leading order correction as:
\begin{eqnarray}
v_n^{(1)}=\frac{1}{2} \left( \frac{\cos(\theta_{n+1}-\theta_n)}{
\sqrt{g(n+1)}} + \frac{\cos(\theta_{n-1}-\theta_n)}{
\sqrt{g(n-1)}} \right) e^{i \theta_n}
\label{ddnls8}
\end{eqnarray}
when for the $n$-th site both of its neighbors are excited; when only
one neighbor is excited, then only the corresponding term is 
present in Eq.~(\ref{ddnls8}). Notice that in this expression
and hereafter for simplicity (and without loss of generality),
we will set $\Lambda=1$. Using this expression the diagonal elements
of the $N \times N$ matrix arising in the numerator of Eq.~(\ref{ddnls7})
${\cal M}=\langle b, {\cal L}_- b \rangle$
are found to be:
\begin{eqnarray}
{\cal M}_{n,n}
= \sqrt{g(n)} \left( \frac{\cos(\theta_{n+1}-\theta_n)}{
\sqrt{g(n+1)}} + \frac{\cos(\theta_{n-1}-\theta_n)}{
\sqrt{g(n-1)}} \right)
\label{ddnls9}
\end{eqnarray}
(again, if both neighbors are excited). On the other hand, the
off diagonal contributions remain the same as in~\cite{pkf},
namely
\begin{eqnarray}
{\cal M}_{n,n \pm 1} =  - \cos(\theta_{n \pm 1} -\theta_n).
\label{ddnls10}
\end{eqnarray}

Once the eigenvalues $\gamma$ of ${\cal M}$ are calculated, then the 
eigenvalues of the full problem bifurcating from $0$
are given as $\lambda= \pm \sqrt{-2 \epsilon \gamma}$.
Let us give some explicit examples. In the case of $N=2$
excited sites, the relevant matrix 
\begin{eqnarray}
{\cal M}= \left( \begin{array}{cc}
\sqrt{\frac{g(n)}{g(n+1)}} & -1 \\
-1  &  \sqrt{\frac{g(n+1)}{g(n)}} \\
    \end{array} \right) \cos(\theta_{n+1}-\theta_n).
\label{ddnls11}
\end{eqnarray}
This leads to $\gamma=0$ and $\gamma=(\sqrt{\frac{g(n)}{g(n+1)}}
+ \sqrt{\frac{g(n+1)}{g(n)}}) \cos(\theta_{n+1}-\theta_n)$.
It is particularly interesting to note that in this setting
$(\sqrt{\frac{g(n)}{g(n+1)}}
+ \sqrt{\frac{g(n+1)}{g(n)}}) \geq 2$, with the latter value
being the homogeneous limit case of $g(n)=1$ (i.e., of all sites
bearing an equal nonlinearity prefactor). This effectively implies
that the inhomogeneous solution will always be more
prone to instability. In the out-of-phase case of
$\cos(\theta_{n+1}-\theta_n)=-1$, this will be because
of a real eigenvalue pair which is larger in magnitude
in the inhomogeneous case. On the other hand, 
in the in-phase case of $\cos(\theta_{n+1}-\theta_n)=1$,
the eigenvalue pair will be imaginary (again
larger in magnitude for the inhomogeneous case) and will
start growing along the imaginary axis as $\epsilon$
is increased. This, in turn,
given (as in the case of~\cite{pkf}; see the relevant
discussion therein) the negative signature
of the relevant eigenvalue, will eventually lead to an instability 
as $\epsilon$ increases, upon the collision of this
eigenvalue with the continuous spectrum; see also below the detailed
discussion associated with Fig.~\ref{kps_fig1}.
Based on the above discussion, we expect the inhomogeneous case to be
more prone to instability than its homogeneous counterpart.

In the case of $N=3$ excited sites, the resulting 
$3 \times 3$ reduced matrix is of the form:
\begin{eqnarray}
{\cal M}= \left( \begin{array}{ccc}
\sqrt{\frac{g(n-1)}{g(n)}} \cos(\theta_{n-1}-\theta_n) & -\cos(\theta_{n-1}-\theta_n) & 0 \\
-\cos(\theta_{n-1}-\theta_n)  &  \sqrt{\frac{g(n)}{g(n-1)}} \cos(\theta_{n-1}-\theta_n) + \sqrt{\frac{g(n)}{g(n+1)}} \cos(\theta_{n+1}-\theta_n) 
& - \cos(\theta_{n+1}-\theta_n)  \\
0 & -\cos(\theta_{n+ 1}-\theta_n) & \sqrt{\frac{g(n+1)}{g(n)}} \cos(\theta_{n+1}-\theta_n) \\
    \end{array} \right).
\label{ddnls12}
\end{eqnarray}
In this case too, one can find explicitly the eigenvalues of the matrix,
although the resulting expression is far more cumbersome. More specifically,
setting $a=\sqrt{\frac{g(n-1)}{g(n)}}$ and $b=\sqrt{\frac{g(n+1)}{g(n)}}$,
in addition to $0$, the other two resulting eigenvalues $\gamma$ in this
case are: $\gamma=(2 a b)^{(-1)} ((1+a^2) \cos(\theta_{n-1}-\theta_n)
+ a \cos(\theta_{n+ 1}-\theta_n) \pm \sqrt{-4 a b(b^2 + a^2 (1+b^2))
\cos(\theta_{n-1}-\theta_n) \cos(\theta_{n+ 1}-\theta_n) + ((1+a^2) 
b \cos(\theta_{n-1}-\theta_n) + a (1+b^2) \cos(\theta_{n+ 1}-\theta_n))^2})$.
In this case, it is less straightforward to provide a general
statement about the comparison to the homogeneous state of $g(n)=1$.
Nevertheless, we would like to note that in the case 
considered here also, the magnitudes of the eigenvalues are 
larger in comparison with the homogeneous case of $g(n)=1$ $\forall n$
(for the corresponding
setting of three-site-excitations within the latter). This, in turn,
provides a stronger (larger growth rate) instability --again
in comparison with the homogeneous  $g(n)=1$ $\forall n$ case-- 
for
configurations with at least an out-of-phase pair of adjacent sites
and an instability arising for smaller values of $\epsilon$ in the
case of (all) excited sites bearing an in-phase structure with respect
to their neighbors.

Lastly, although we give no quantitative information about that
case, it is relevant to add a brief note regarding the extended 
excitation. In the latter case, it is important to point
out that {\it all} eigenvalues are at $0$ in the AC limit.
Hence, the size of the matrix ${\cal M}$ in this case is 
comparable to the domain size and hence it is less straightforward
to characterize the relevant eigenvalues. On the other hand, it
is especially relevant to report that since the pulse is decaying
in the case of a potential growing at infinity, the corresponding
effective potential will be unbounded as $n \rightarrow \infty$
in that case. This, in turn, leads the eigenvalues to bifurcate
from the origin of the spectral plane with $\lambda=0$ giving
rise to a point (rather than continuous) spectrum. 
We now turn to the numerical examination of the relevant findings.

\begin{table}[htdp]
\caption{Existence intervals for the solutions considered herein.
The first column labels the branches, while the second provides
their profile form near the Anti-Continuum limit. The third
column provides the end point of their termination (for branch
G for the coupling values considered herein, no such end point
was identified, hence the N/A symbolism). Finally, the fourth
column illustrates the fate of the branches i.e., the nature
of the bifurcation and with which branch they  collide. It should
be highlighted that as the bifurcation is approached, the  
{\it shapes} of the two (or more) colliding branches become
fairly similar i.e., the deformation of branch B bears resemblance
to branch E, and so on. In the case of branch D the collision 
occurs with the more extended branch $(0,\dots,0,-\sqrt{\frac{1}{g(-2)}},\sqrt{\frac{1}{g(-1)}},\sqrt{\frac{1}{g(0)}},\sqrt{\frac{1}{g(1)}},-\sqrt{\frac{1}{g(2)}},0,\dots0)$.} 
\centering 
\begin{tabular}{c c c c } 
\hline\hline 
Label  & Structure & Terminal Point & Endpoint Bifurcation \\ [0.5ex] 
\hline
A & $(0,\dots,0,\sqrt{\frac{1}{g(0)}},0,\dots0)$ & $\epsilon=0.095$ & 
Double Pitchfork with C, F \\
B &  $(0,\dots,0,\sqrt{\frac{1}{g(0)}},\sqrt{\frac{1}{g(1)}},0,\dots0)$ & $\epsilon=0.091$ & 
Saddle-Center with E \\
C & $(0,\dots,0,\sqrt{\frac{1}{g(0)}},-\sqrt{\frac{1}{g(1)}},0,\dots0)$ & $\epsilon=0.095$ & 
Double Pitchfork with A, F \\
D & $(0,\dots,0,\sqrt{\frac{1}{g(-1)}},\sqrt{\frac{1}{g(0)}},\sqrt{\frac{1}{g(1)}},0,\dots0)$  & $\epsilon=0.121$ & 
Saddle-Center \\
E &  $(0,\dots,0,-\sqrt{\frac{1}{g(-1)}},\sqrt{\frac{1}{g(0)}},\sqrt{\frac{1}{g(1)}},0,\dots0)$  & $\epsilon=0.091$ & 
Saddle-Center with B \\
F & $(0,\dots,0,-\sqrt{\frac{1}{g(-1)}},\sqrt{\frac{1}{g(0)}},-\sqrt{\frac{1}{g(1)}},0,\dots0)$  & $\epsilon=0.095$ & 
Double Pitchfork with A, C \\
G & $v_n= \sqrt{1/g(n)}$ & N/A & 
N/A \\
\hline
\end{tabular}
\label{t1}
\end{table}%

\section{Numerical Computations}

For our concrete numerical example, we will use a power law
growth of the nonlinear prefactor in the
form $g(n)=1 + 10 n^2$. This choice is made purely for purposes
of illustration, as the above general theory, in principle, enables
the computation of the relevant states and their linearization eigenvalues
for arbitrary forms of $g(n)$.
Arguably, the most fundamental branch of numerical solutions is the
one with a single site excitation  
$(0,\dots,0,\sqrt{\frac{1}{g(0)}},0,\dots0)$ at the AC limit
(of $\epsilon=0$). We will only touch
upon this branch of solutions briefly at present and return
to it, as we consider the bifurcations of more complex branches
of solutions. In the homogeneous case where $g(n)$ is constant,
this branch would persist to large values of the coupling parameter
forming the discrete analogue of the gap solitary wave in this model.
Here, however, this is no longer true. Our computations show that
this branch {\it terminates} around $\epsilon=0.095$, by colliding
with other solution branches as illustrated below. Interestingly,
at the stability level, this branch is stable throughout its interval
of existence. As this termination 
limit is approached, two eigenvalue pairs that
bifurcate off of the continuous spectrum for $\epsilon > 0.045$
approach the spectral plane origin, hitting it at
the critical point, a point indicative of the complex bifurcation
scenario that will be further elaborated below.

\begin{figure}[!ht]
\begin{center}
\includegraphics[width=0.45\textwidth]{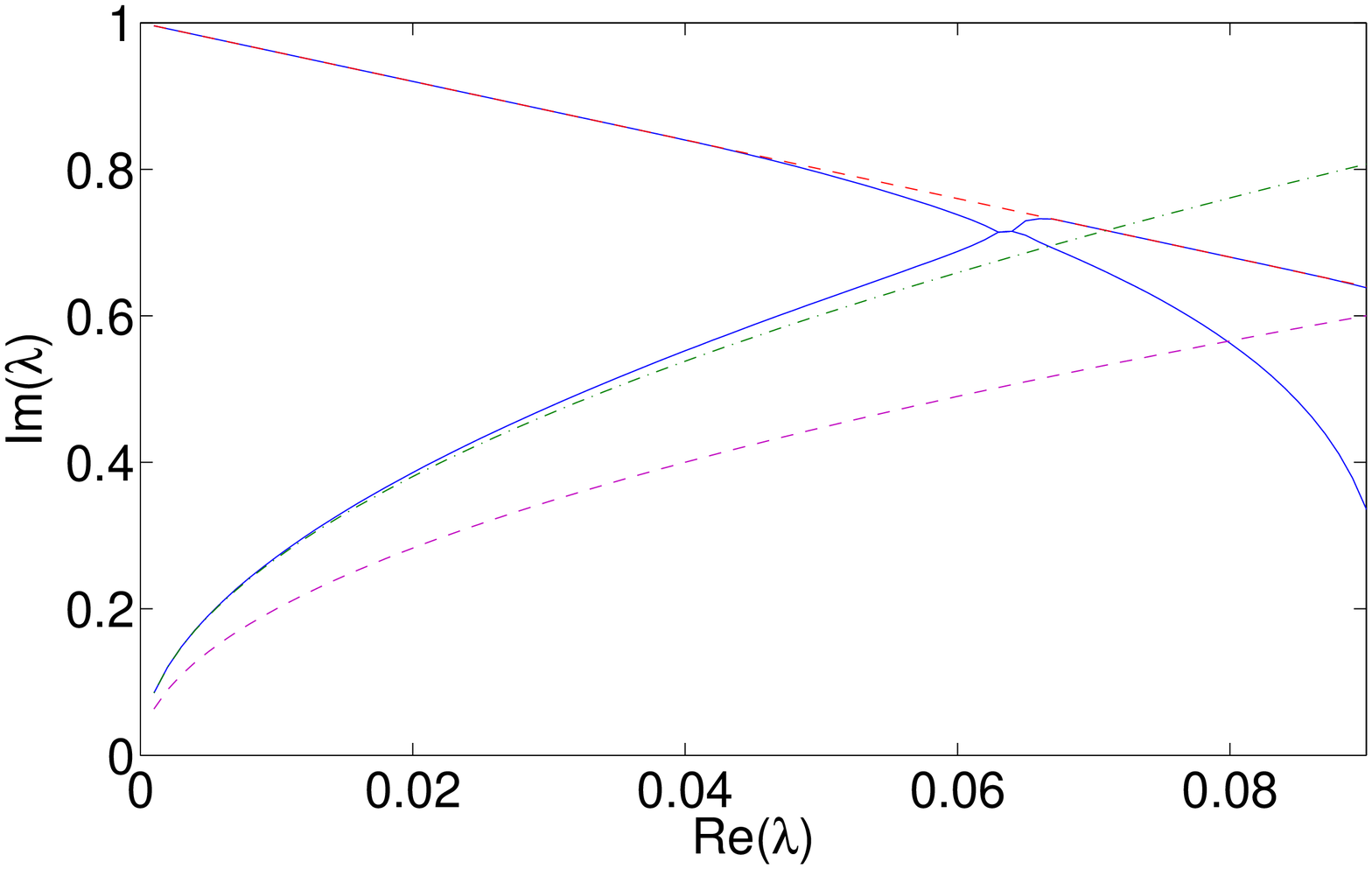}
\includegraphics[width=0.45\textwidth]{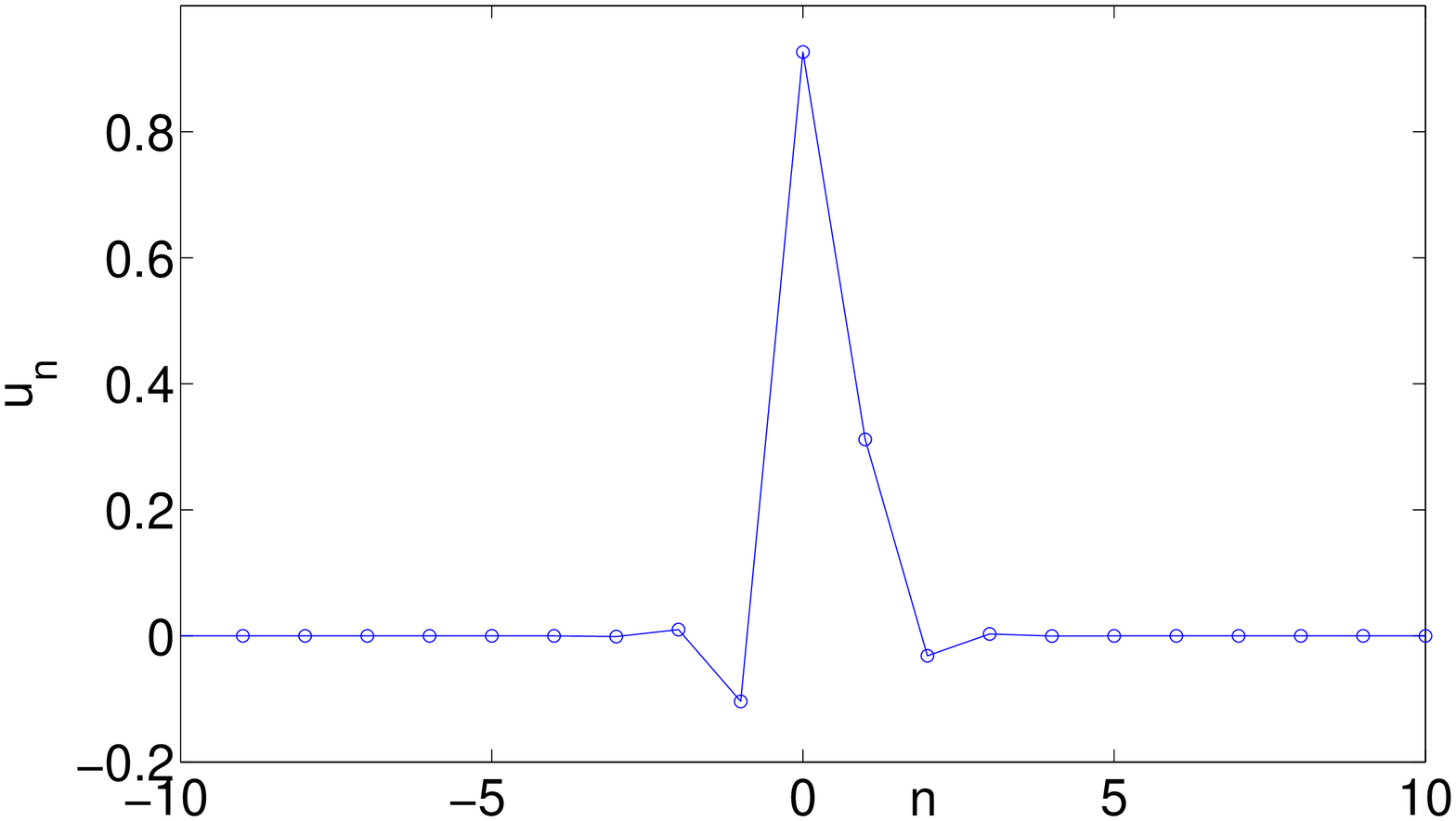}
\end{center}
\begin{center}
\includegraphics[width=0.45\textwidth]{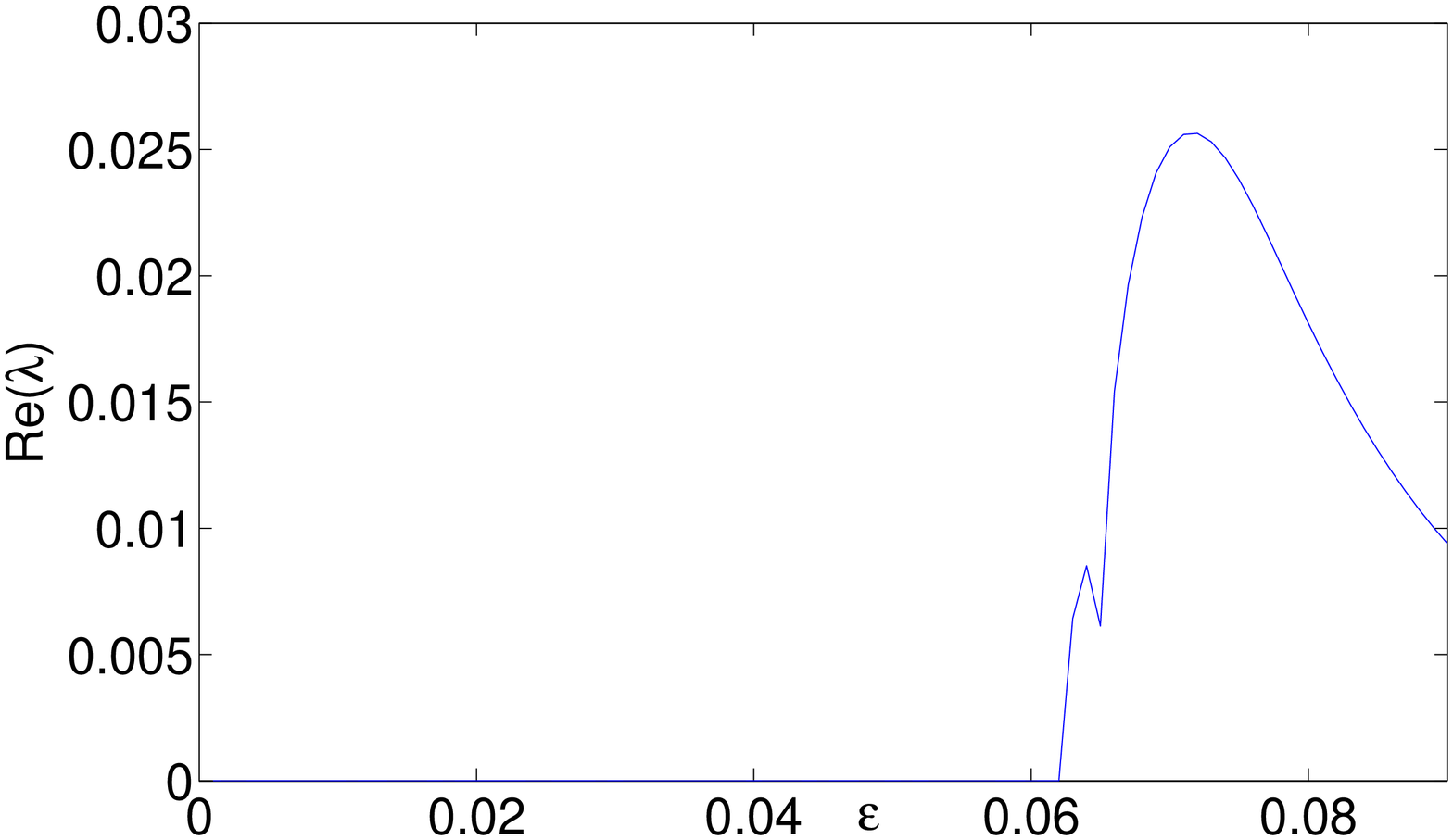}
\includegraphics[width=0.45\textwidth]{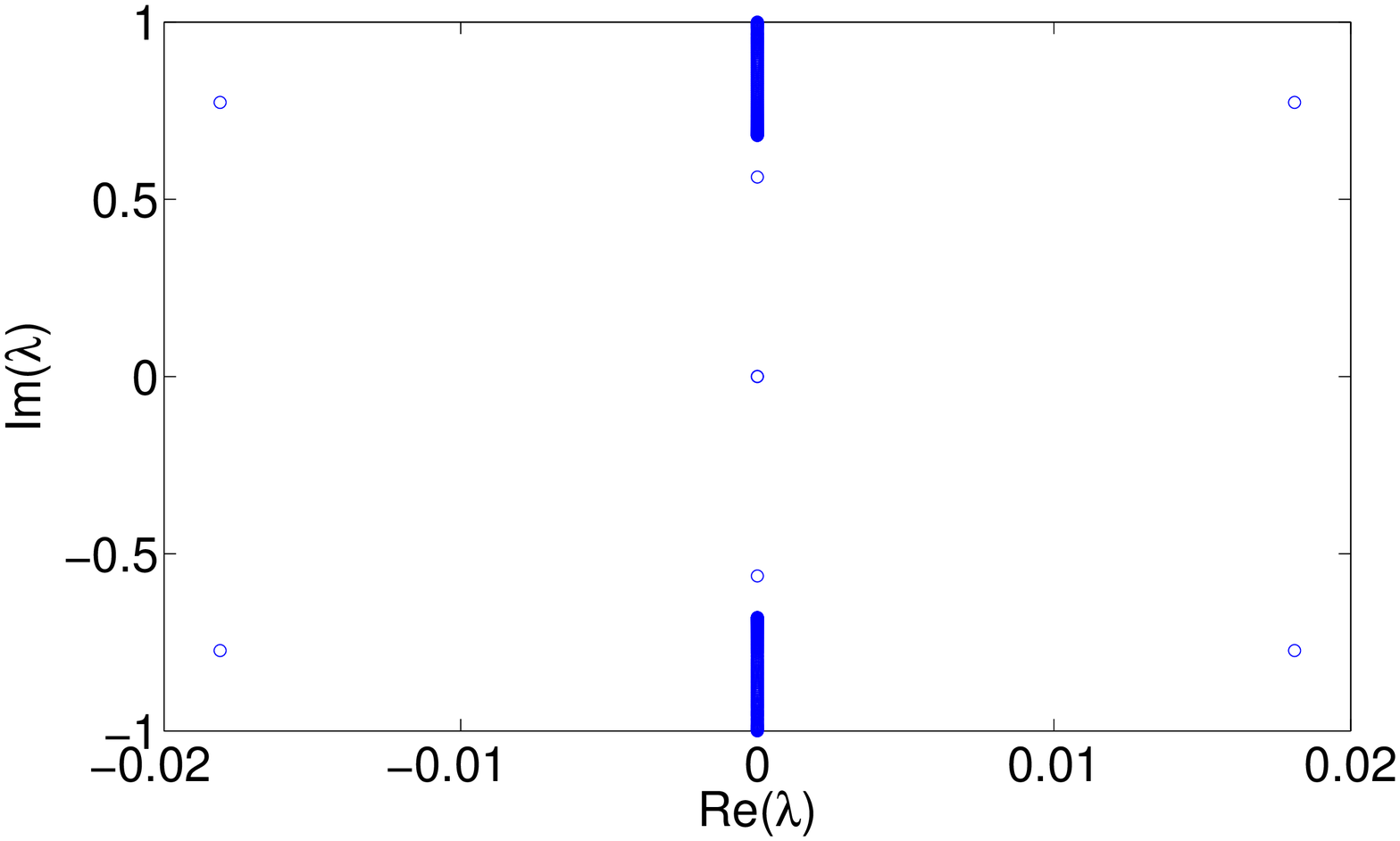}
\end{center}
\caption{The left panels of the figure show the principal 
eigenvalues (top panel: imaginary part; bottom panel: real part)
associated with the branch of solutions $(0,\dots,0,\sqrt{\frac{1}{g(0)}},\sqrt{\frac{1}{g(1)}},0,\dots0)$, while the right panels show a typical profile
of the solution $u_n$ and of its associated spectral plane 
$(Re(\lambda),Im(\lambda))$ of the eigenvalues $\lambda=Re(\lambda)
+ i Im(\lambda)$ for $\epsilon=0.08$. In the top left the numerical eigenvalues
stemming from the origin and from the band edge of the continuous spectrum
are shown by a (blue) solid line. The band edge $Im(\lambda)=1-4 \epsilon$
is shown by a (red) dashed line. The theoretically predicted approximation
of the pair bifurcating from the origin is shown by the (green) dash-dotted
line, while for comparison the (lower) prediction of the homogeneous limit
of $Im(\lambda)=2 \sqrt{\epsilon}$ is also shown (in magenta dashed
line). The bottom
left illustrates that the collision of the pair from the origin with 
eigenvalues at or bifurcating from the continuous spectrum yield
an instability for $\epsilon > 0.062$ (see also the detailed discussion
in the text).}
\label{kps_fig1}
\end{figure}

We now turn to two-site solutions. We start, in particular, by considering
the branch $(0,\dots,0,\sqrt{\frac{1}{g(0)}},\sqrt{\frac{1}{g(1)}},0,\dots0)$
that is shown in Fig.~\ref{kps_fig1}. A typical solution of this
sort is shown in the top right panel of the figure for $\epsilon=0.08$,
for which the corresponding example of the spectral plane 
$(Re(\lambda),Im(\lambda))$ of the eigenvalues $\lambda=Re(\lambda)
+ i Im(\lambda)$ is shown in the bottom right panel. The left panels
in the figure illustrate the evolution of the dominant imaginary 
(top) and real (bottom) parts. 
We can see that as expected from the theory of the previous
section and the eigenvalues of the matrix (\ref{ddnls11}), 
out of the two pairs at the spectral plane origin at $\epsilon=0$
(due to the two excited sites), one will bifurcate along the
imaginary axis (blue parabolic line emanating from $0$ in the top
left panel) being well described by the theoretical prediction
$\lambda \approx \pm 2.69 \sqrt{\epsilon} i$ (green dash-dotted line).
The line of $\lambda=\pm 2 \sqrt{\epsilon} i$ shows the
corresponding eigenvalue pair for the homogeneously nonlinear
case of $g(n)=1$ 
for comparison, clearly illustrating the significant deviation 
of the present inhomogeneous prediction. The lower edge
of the continuous spectrum
is shown by the red-dashed line. Interestingly an eigenvalue
pair bifurcating from the latter (for $\epsilon > 0.04$) collides with
the pair stemming from the origin around $\epsilon=0.062$ destabilizing 
the branch. Although this quartet briefly separates into two pairs
again shortly thereafter, as $\epsilon$ is (slightly) further increased
the pair coming from the origin collides with the band edge of the 
continuous spectrum ensuring a quartet (oscillatory) instability for
all larger values of $\epsilon$ for which the branch exists. 
The branch appears to terminate around 
 $\epsilon=0.091$ due to its collision with
$(0,\dots,0,-\sqrt{\frac{1}{g(-1)}},\sqrt{\frac{1}{g(0)}},\sqrt{\frac{1}{g(1)}},0,\dots0)$, as we will also see in what follows.

\begin{figure}[!ht]
\begin{center}
\includegraphics[width=0.45\textwidth]{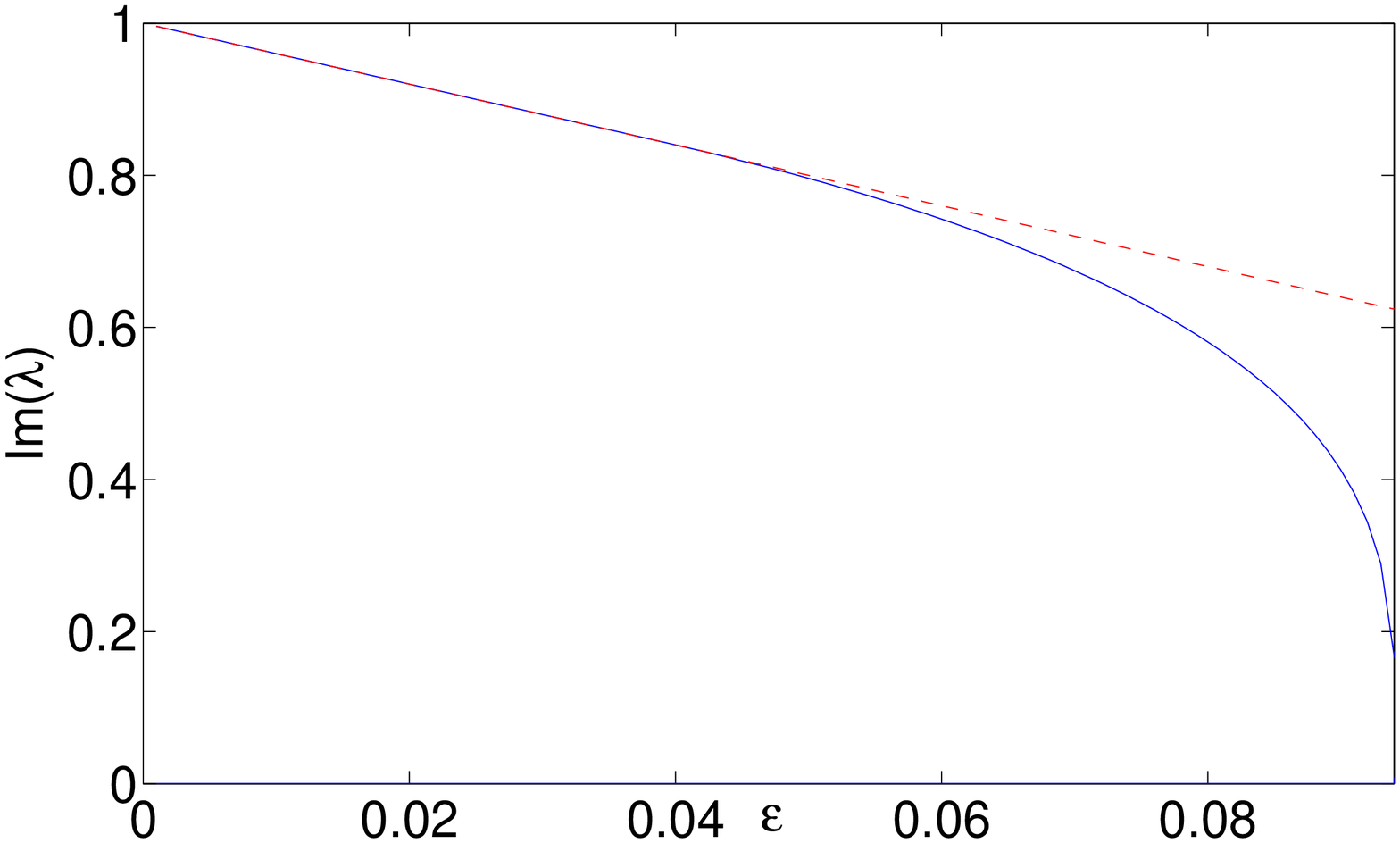}
\includegraphics[width=0.45\textwidth]{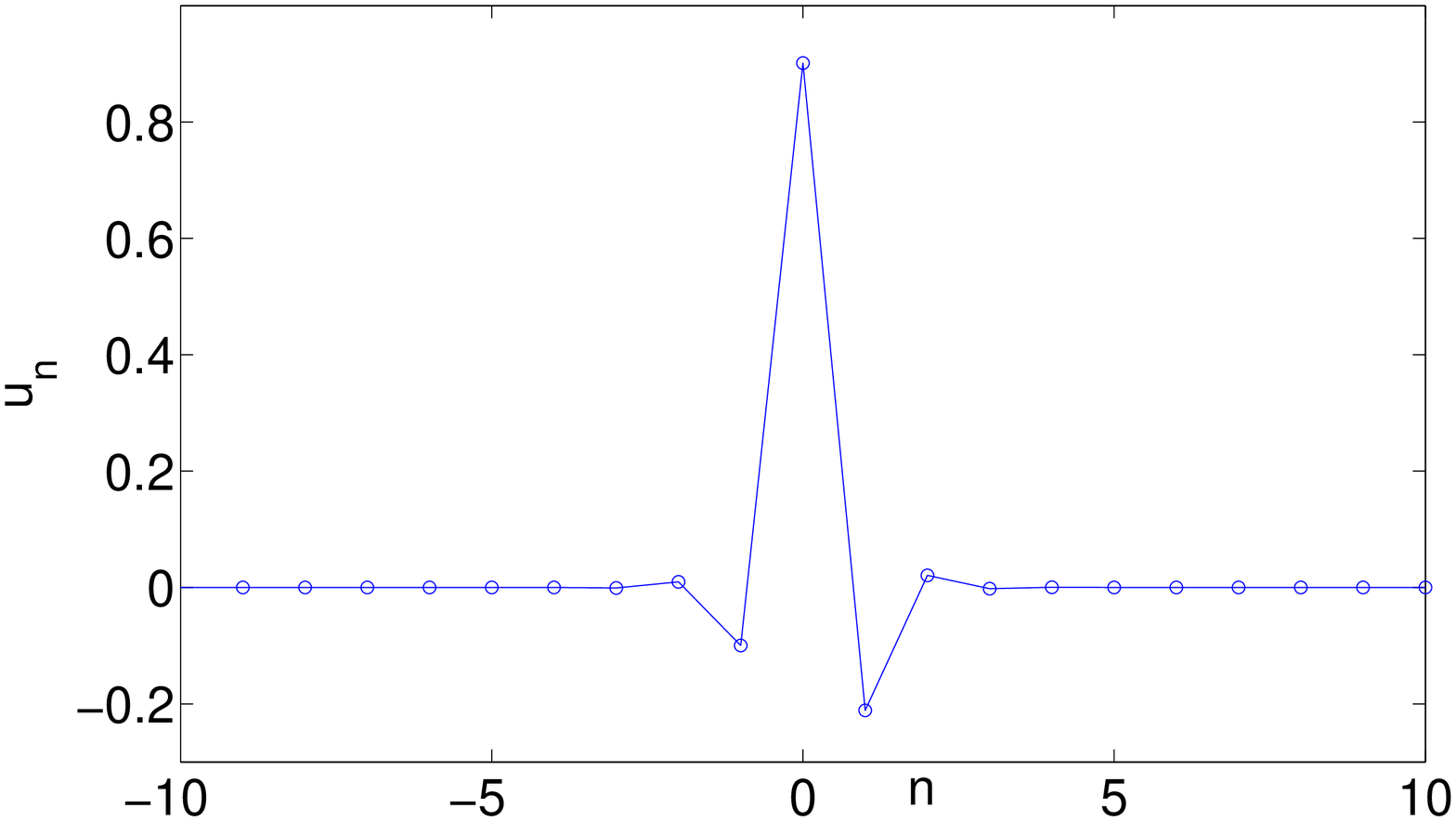}
\end{center}
\begin{center}
\includegraphics[width=0.45\textwidth]{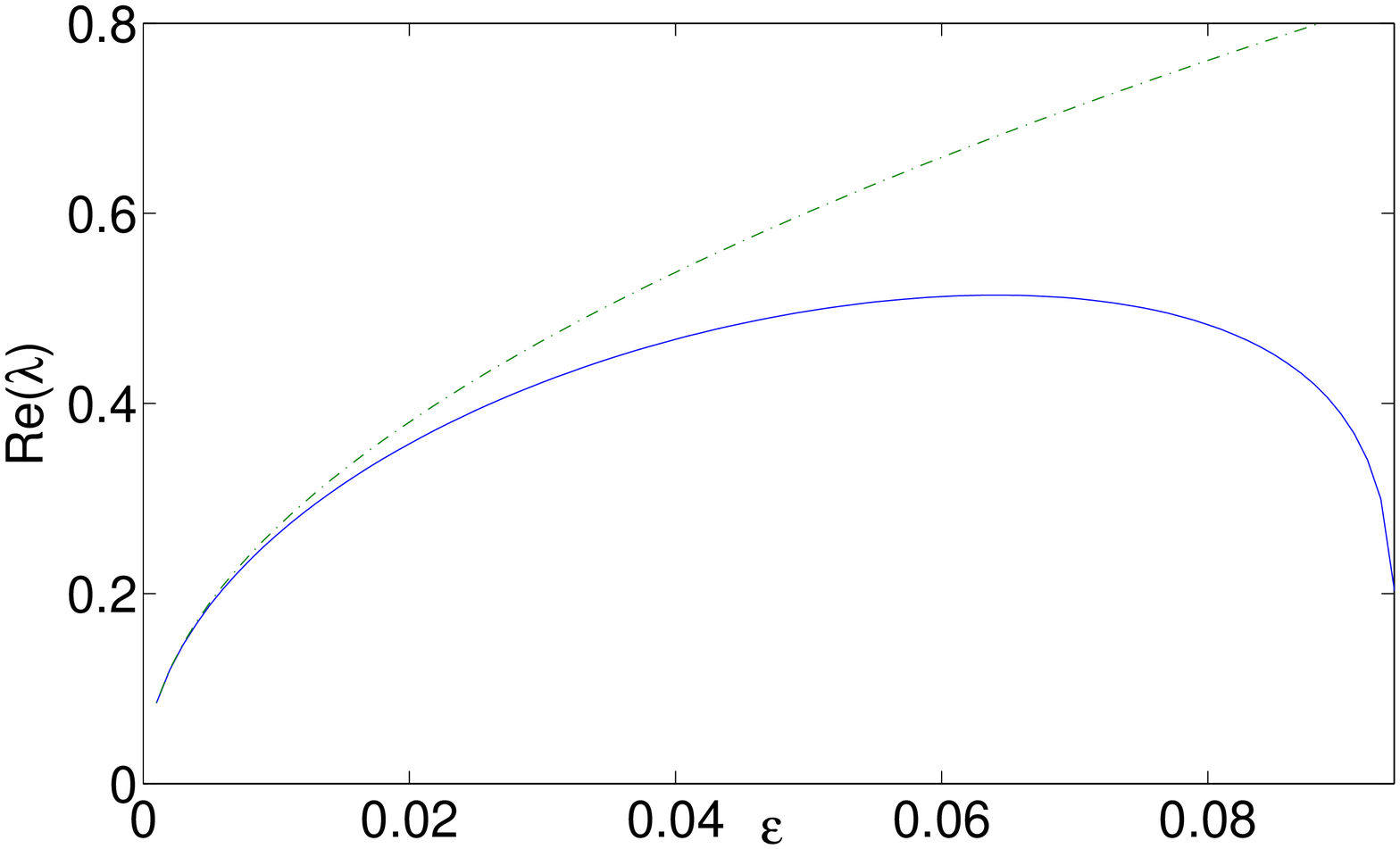}
\includegraphics[width=0.45\textwidth]{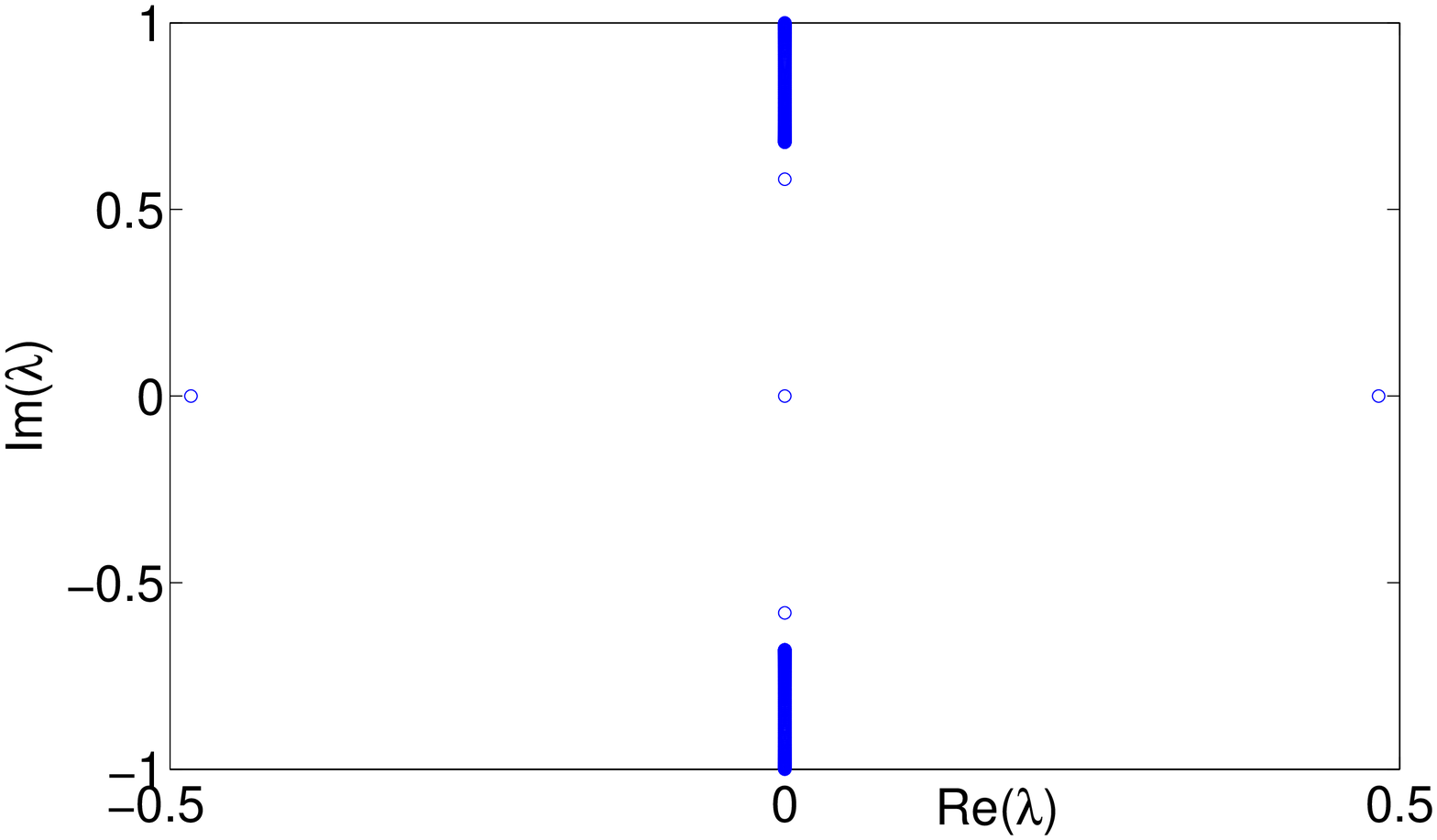}
\end{center}
\caption{Similar to Fig.~\ref{kps_fig1}, but now for the out-of-phase
two-site branch with profile
$(0,\dots,0,\sqrt{\frac{1}{g(0)}},-\sqrt{\frac{1}{g(1)}},0,\dots0)$ at
the AC-limit.
Here a pair bifurcating from the continuous spectrum and the real
pair bifurcating from the origin (returning to it after an excursion
along the real line) collide at the disappearance threshold of the branch
$\epsilon=0.095$. Again the top right and the bottom right panels
illustrate, respectively, the (asymmetric) profile of the solution
and its spectral plane for $\epsilon=0.08$.}
\label{kps_fig2}
\end{figure}

As the second example of a two-site excitation branch, we illustrate
in Fig.~\ref{kps_fig2} the out-of-phase case of
$(0,\dots,0,\sqrt{\frac{1}{g(0)}},-\sqrt{\frac{1}{g(1)}},0,\dots0)$.
In this case the bifurcation from the origin occurs along the
real (rather than the imaginary) axis, leading to an {\it immediate}
instability of the solution. It is relevant to point out here the
differences of this case from the corresponding focusing case; see 
e.g. for a relevant discussion~\cite{hadisus} and also for a 
review~\cite{book}. In the focusing case, multi-site in-phase excitations
are immediately unstable with eigenvalues bifurcating from the origin
on the real line, while ones with all adjacent neighbors out-of-phase
are linearly stable, at least close to the AC limit. The situation
is reversed in the defocusing realm, as illustrated in this and
the previous example, a pattern that will also be followed in the
three-site excitations below. The theoretical prediction 
for the real pair is $\lambda \approx 2.69 \sqrt{\epsilon}$, which 
we can see as a good approximation to leading order for small $\epsilon$,
but one that progressively fails as higher orders take over for larger
$\epsilon$ and force the relevant pair back to the origin where it
collides with an imaginary pair at $\epsilon=0.095$. It may now
be becoming clearer, both from the eigenvalue pattern and
associated branch extinction collision point, as well as perhaps
from the profile of the branch (top right of Fig.~\ref{kps_fig2})
and its progressive ``symmetrization'' as the critical point is
approached that this branch is involved together with the single
site branch in the complex bifurcation further elaborated below.


\begin{figure}[!ht]
\begin{center}
\includegraphics[width=0.45\textwidth]{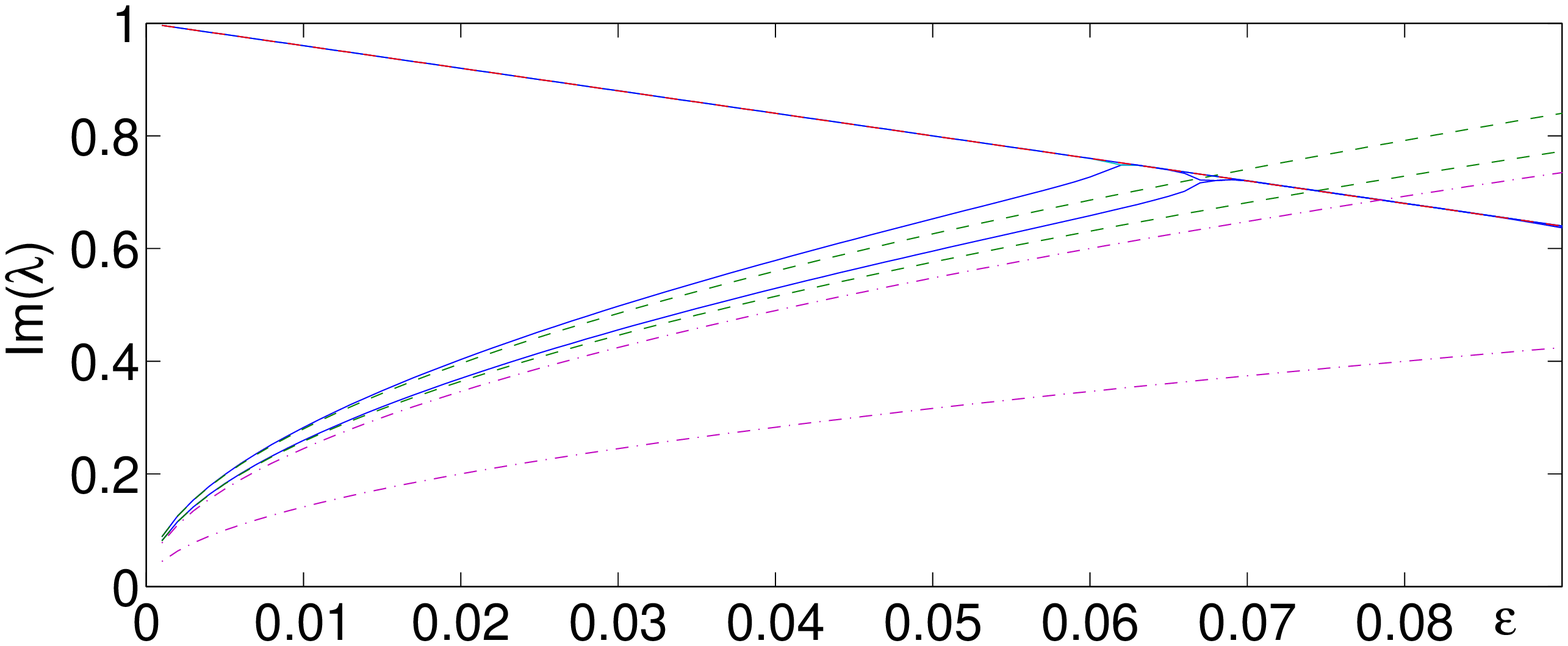}
\includegraphics[width=0.45\textwidth]{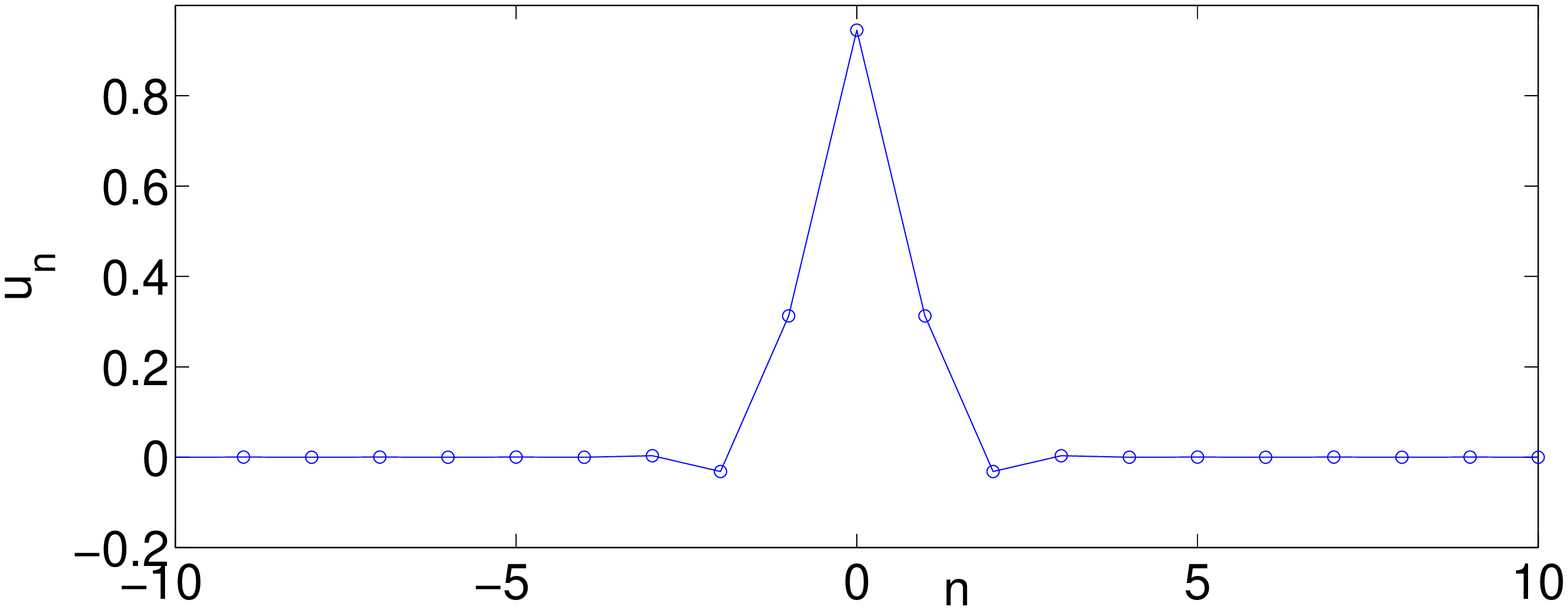}
\end{center}
\begin{center}
\includegraphics[width=0.45\textwidth]{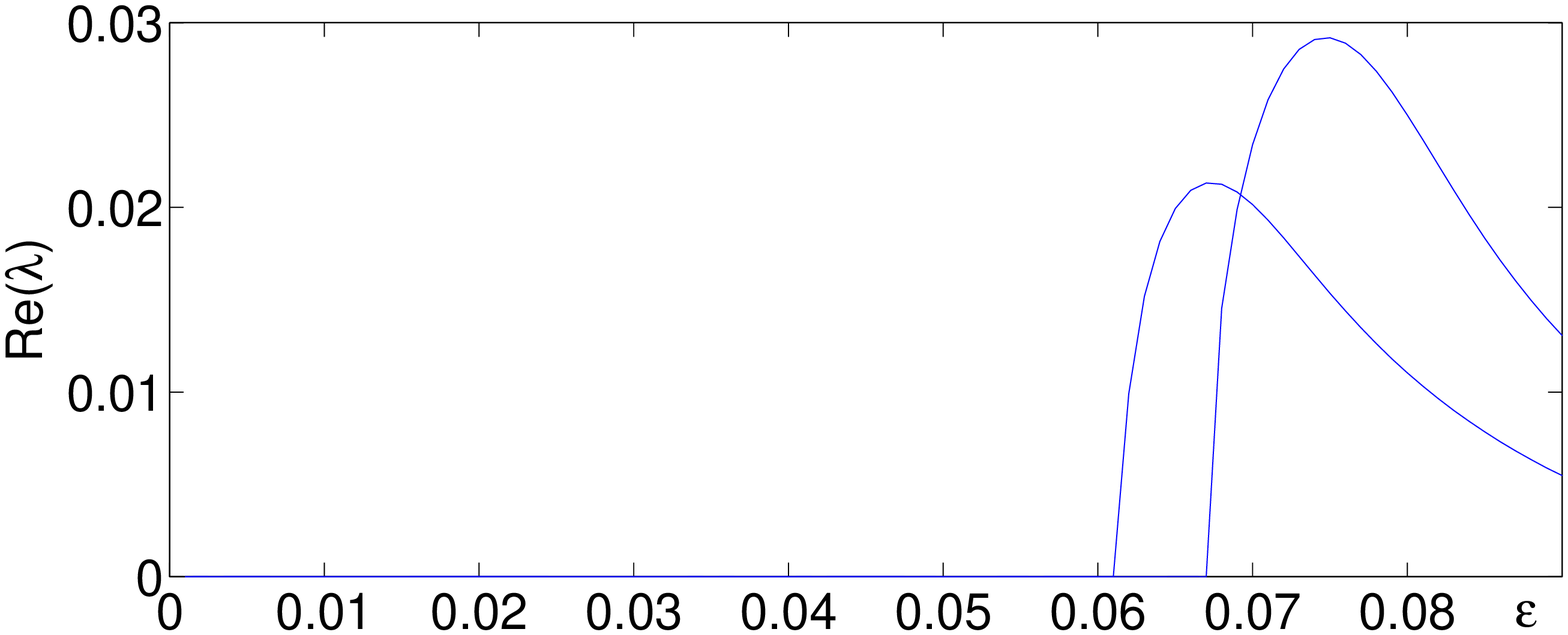}
\includegraphics[width=0.45\textwidth]{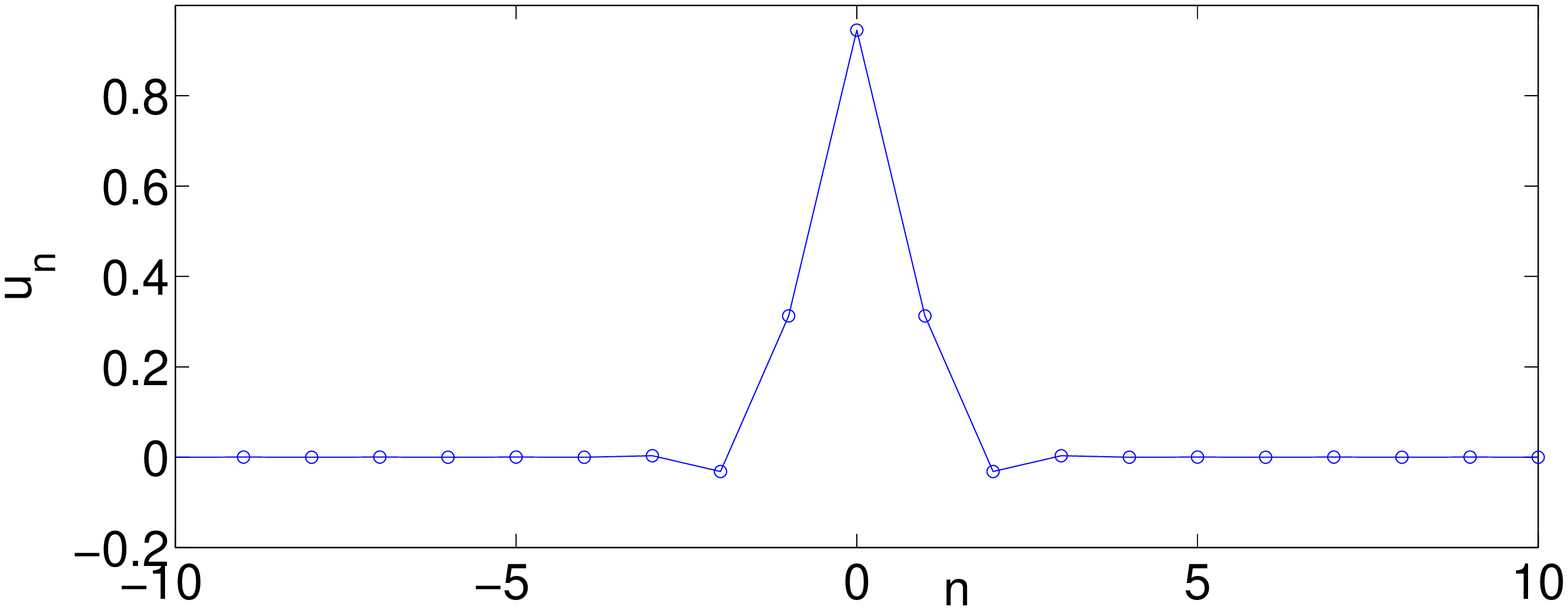}
\end{center}
\caption{Same as with the previous branches, 
but now for the solution branch with the profile
$(0,\dots,0,\sqrt{\frac{1}{g(-1)}},\sqrt{\frac{1}{g(0)}},\sqrt{\frac{1}{g(1)}},0,\dots0)$ at the AC limit.
The main difference
here is that there are two imaginary eigenvalue pairs bifurcating from the
origin, and two associated quartets of eigenvalues arising beyond
$\epsilon=0.061$ and $\epsilon=0.067$, respectively.}
\label{kps_fig3}
\end{figure}

Next we turn to three-site excitations of which (modulo permutations,
similarly to~\cite{pkf}) we examine three examples in what
follows. We start with the
in-phase branch 
$(0,\dots,0,\sqrt{\frac{1}{g(-1)}},\sqrt{\frac{1}{g(0)}},\sqrt{\frac{1}{g(1)}},0,\dots0)$. This branch, in full accordance with the theory
of the previous section (cf. Eq.~(\ref{ddnls12})), has
two imaginary eigenvalues bifurcating from the origin
along the imaginary axis (while the third of the AC limit
pairs of $\lambda=0$ due to the three excited sites remains
at $0$, given the phase invariance of the model). 
The corresponding eigenvalue pairs are theoretically predicted
from Eq.~(\ref{ddnls12}) to be
$\lambda=\pm 2.576 \sqrt{\epsilon} i$ and
$\lambda= \pm 2.8 \sqrt{\epsilon} i$ and are found to be in very
good agreement with the numerical findings (cf. the top left
panel of Fig.~\ref{kps_fig3}). Additionally, as in the two-site
case, these eigenvalues are found to be larger
than their corresponding homogeneous limit predictions, shown 
by the magenta dashed lines in the figure. This indeed implies also
that the instability of the branch occurs for a smaller 
value of $\epsilon$ in comparison with the homogeneous limit,
as it arises from the collision of the two imaginary
eigenvalue pairs with the continuous spectrum
of $\lambda=1-4 \epsilon$. In this case, given the two collisions,
two oscillatory instabilities and associated quartets arise
for $\epsilon > 0.061$ and $0.067$, 
respectively. In this case, it is interesting to point out
that the branch terminates for values larger than any of the above
(as well as below) few site excited branches. 
More specifically, it
collides with the
5-site branch $(0,\dots,0,-\sqrt{\frac{1}{g(-2)}},\sqrt{\frac{1}{g(-1)}},\sqrt{\frac{1}{g(0)}},\sqrt{\frac{1}{g(1)}},-\sqrt{\frac{1}{g(2)}},0,\dots0)$
at $\epsilon=0.121$.
This is part of a more general trend that we will also discuss below
in the
context of the extended solutions. In particular, the more extended
a solution is, the larger the critical threshold value for its
termination.

The fourth branch is the ``mixed'' branch with the AC limit form:
$(0,\dots,0,-\sqrt{\frac{1}{g(-1)}},\sqrt{\frac{1}{g(0)}},\sqrt{\frac{1}{g(1)}},0,\dots0)$. This branch has a real and an imaginary pair of
eigenvalues with $\lambda=\pm 2.685 \sqrt{\epsilon}$ and
$\lambda=\pm 2.685 \sqrt{\epsilon} i$, respectively; the
real eigenvalue pairs renders the branch generically unstable
(similarly to its homogeneous counterpart) although an additional
oscillatory instability arises for $\epsilon>0.065$.
We can see that the numerical imaginary eigenvalue is
very accurately predicted by the theory. For the real one, on
the other hand, we again observe the familiar feature of 
good agreement for small $\epsilon$, but then as $\epsilon$
increases, higher orders take over leading to a maximal excursion
along the real line and a return to $0$ around $\epsilon=0.091$
which is the termination point of the branch in a saddle-center
collision with 
$(0,\dots,0,\sqrt{\frac{1}{g(0)}},\sqrt{\frac{1}{g(1)}},0,\dots0)$.

\begin{figure}[!ht]
\begin{center}
\includegraphics[width=0.45\textwidth]{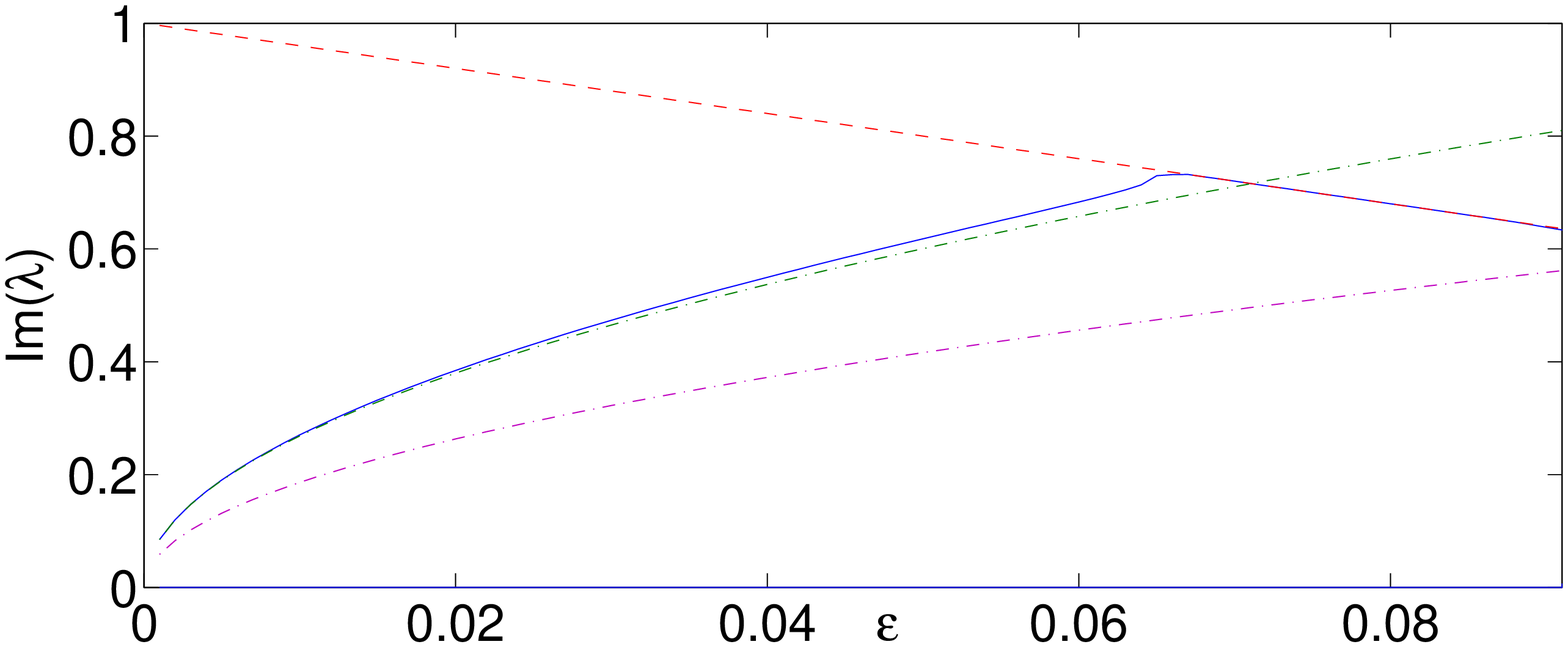}
\includegraphics[width=0.45\textwidth]{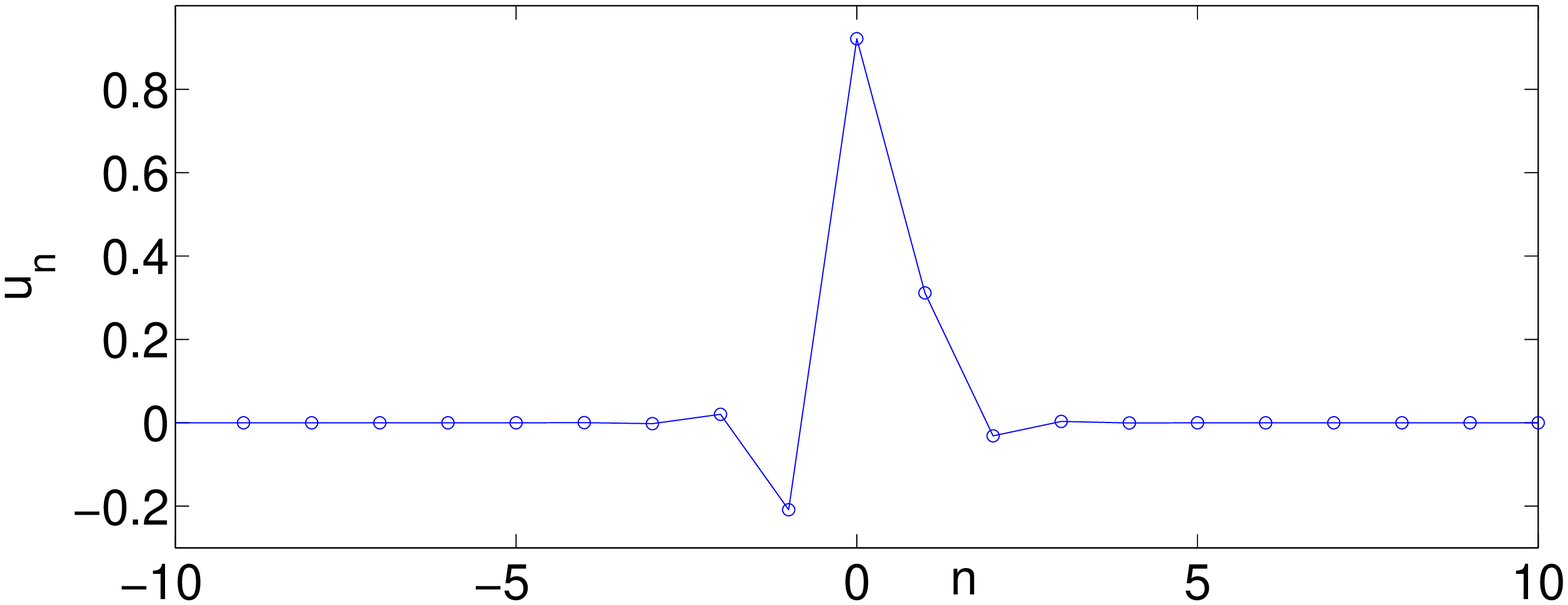}
\end{center}
\begin{center}
\includegraphics[width=0.45\textwidth]{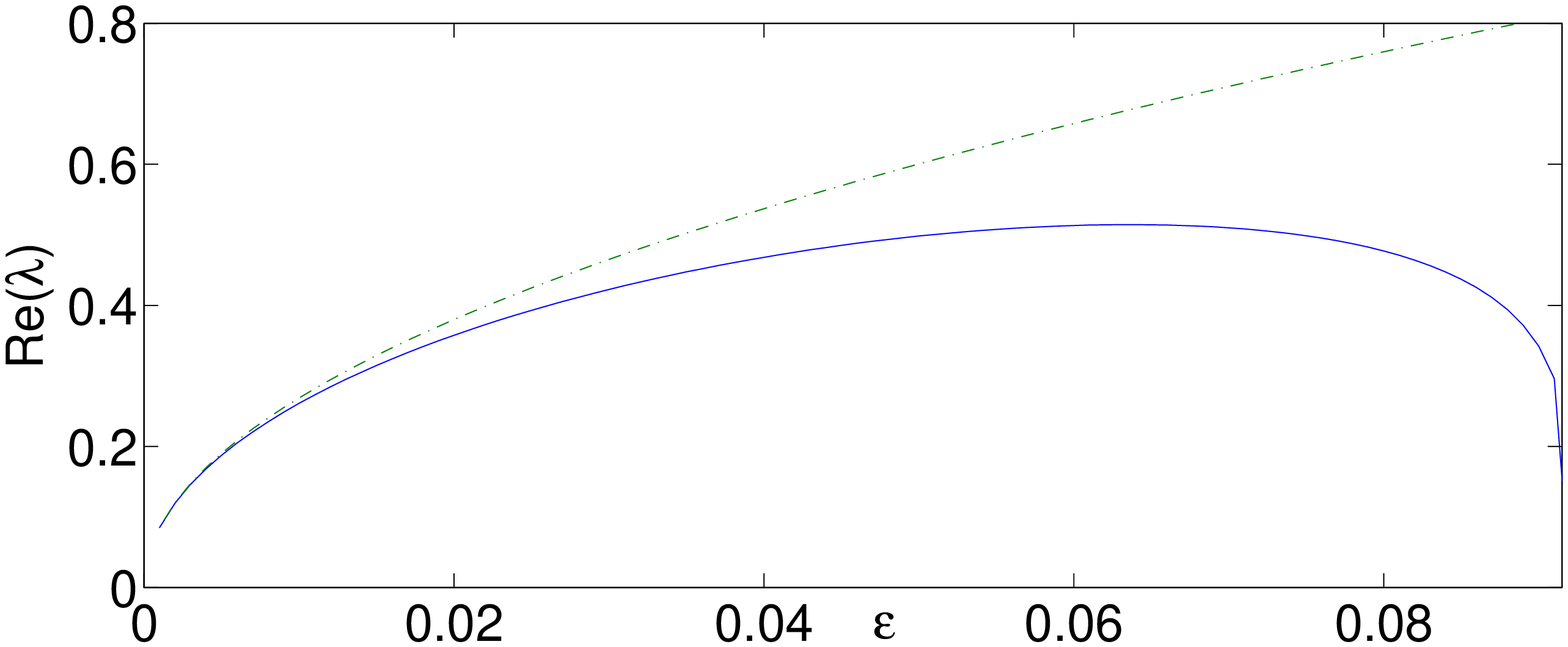}
\includegraphics[width=0.45\textwidth]{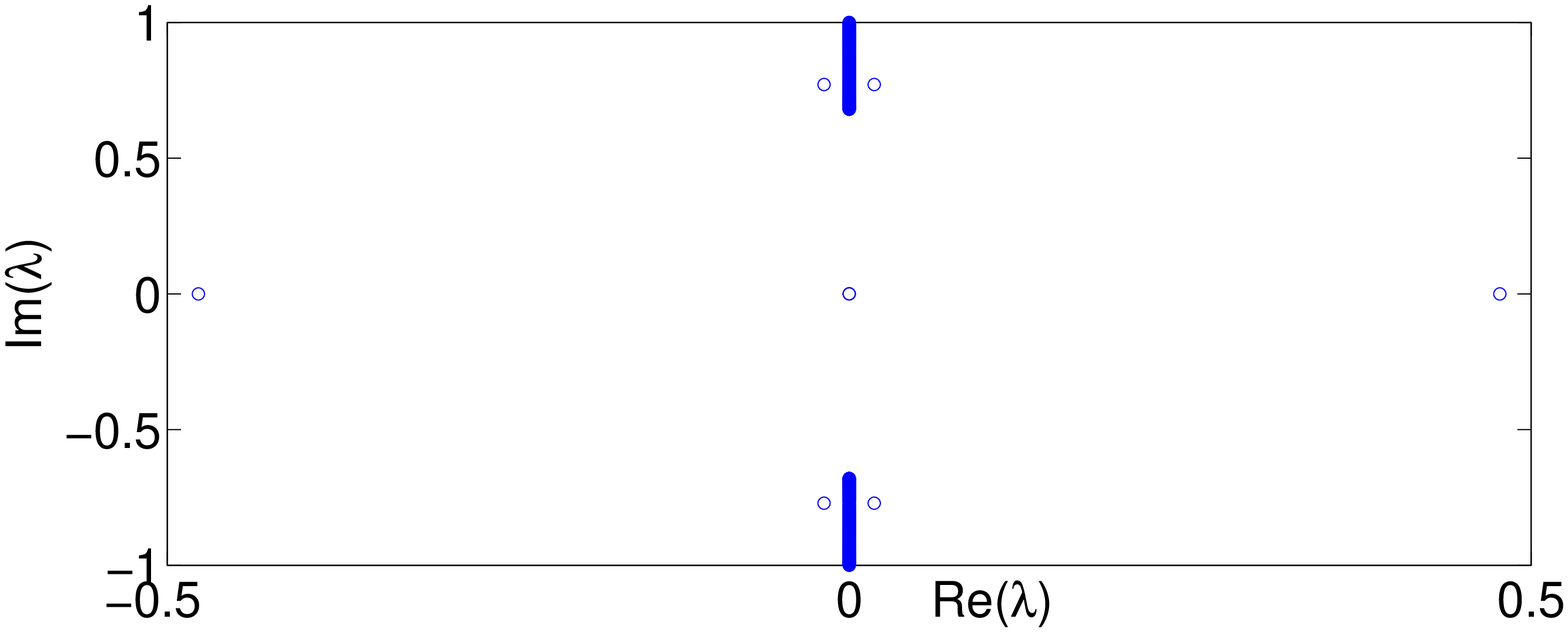}
\end{center}
\caption{Same as for the previous branches, but now for the
``mixed'' solution with profile $(0,\dots,0,-\sqrt{\frac{1}{g(-1)}},\sqrt{\frac{1}{g(0)}},\sqrt{\frac{1}{g(1)}},0,\dots0)$ in the AC-limit. Here, one
of the eigenvalue pairs moving off of the origin for $\epsilon \neq 0$
moves along the real and one along the imaginary axis.}
\label{kps_fig4}
\end{figure}

Finally, from the point of view of few-site excitations, 
we explore the out-of-phase three-site branch
of the form (at the AC-limit) 
$(0,\dots,0,-\sqrt{\frac{1}{g(-1)}},\sqrt{\frac{1}{g(0)}},-\sqrt{\frac{1}{g(1)}},0,\dots0)$. Here the stability matrix is the same as in the
in-phase case, but with an opposite sign, hence the eigenvalue
predictions of the theory of Eq.~(\ref{ddnls12}) are the
same as in the former case, but along the real axis, as opposed
to along the imaginary one. As we have seen multiple times
with real eigenvalues, the predictions are fairly accurate for
small $\epsilon$, but for large values of the parameter, higher
orders take over and lead the pairs to return to the origin.
Here, both pairs return to the origin around $\epsilon=0.095$,
the point of the termination of the branch. This end
point is intriguingly the same as the termination point
of both $(0,\dots,0,\sqrt{\frac{1}{g(0)}},0,\dots0)$ and
(the two mirror image installments)
$(0,\dots,0,\sqrt{\frac{1}{g(0)}},-\sqrt{\frac{1}{g(1)}},0,\dots0)$ 
and
$(0,\dots,0,-\sqrt{\frac{1}{g(-1)}},\sqrt{\frac{1}{g(0)}},0,\dots0)$.
In effect, we see a rather unusual bifurcation scenario here, which
appears as a sort of ``double pitchfork''. Namely, there are two
pairs of eigenvalues involved (hence the ``double'' designation).
For the branch with a single excited site, these eigenvalues both
come from the imaginary side (bifurcating from the continuous spectrum),
while for the three-site out-of-phase branch, they both come from the
side of the real axis. For the asymmetric branches, the two pairs
are split with one on the real and one on the imaginary axis.
Hence, the bifurcation effectively involves a highly symmetric
pair of subcritical pitchforks, ultimately leading to the 
termination of {\it all} 4 associated branches.  

\begin{figure}[!ht]
\begin{center}
\includegraphics[width=0.45\textwidth]{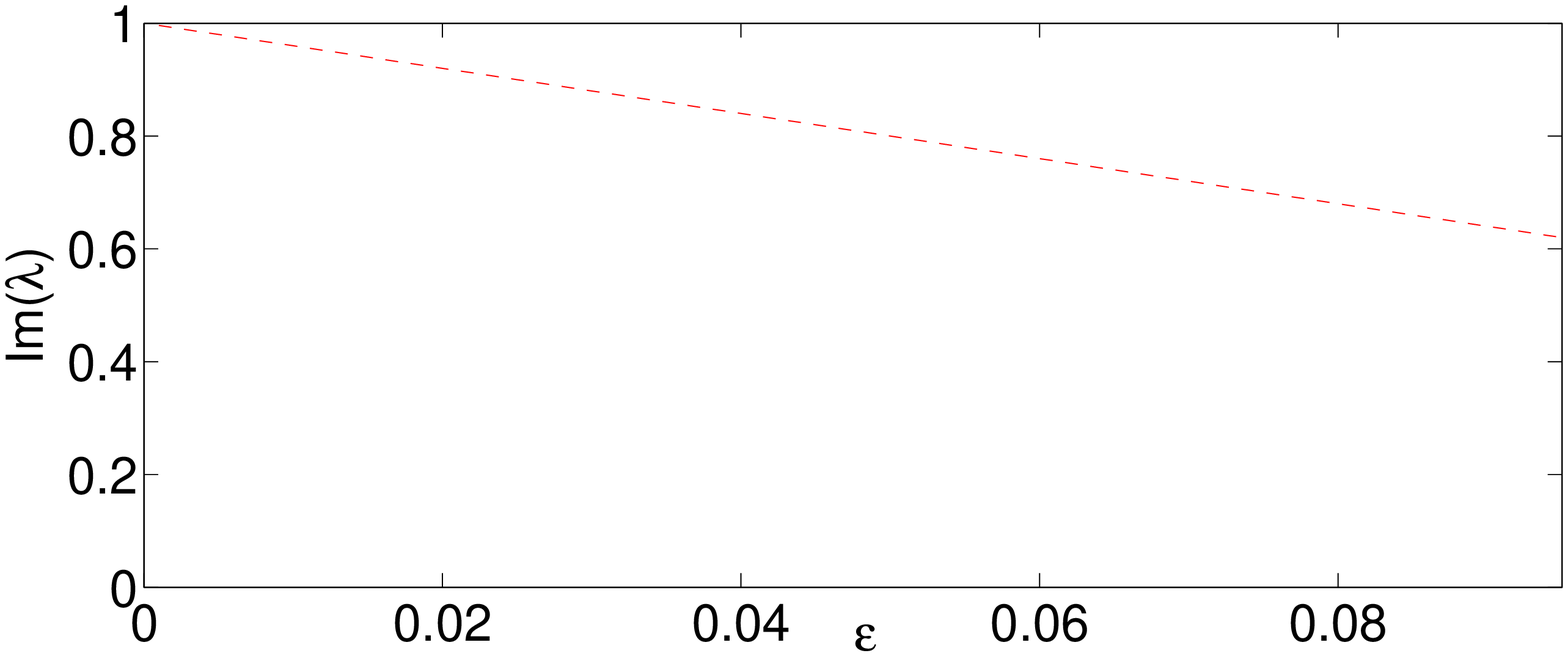}
\includegraphics[width=0.45\textwidth]{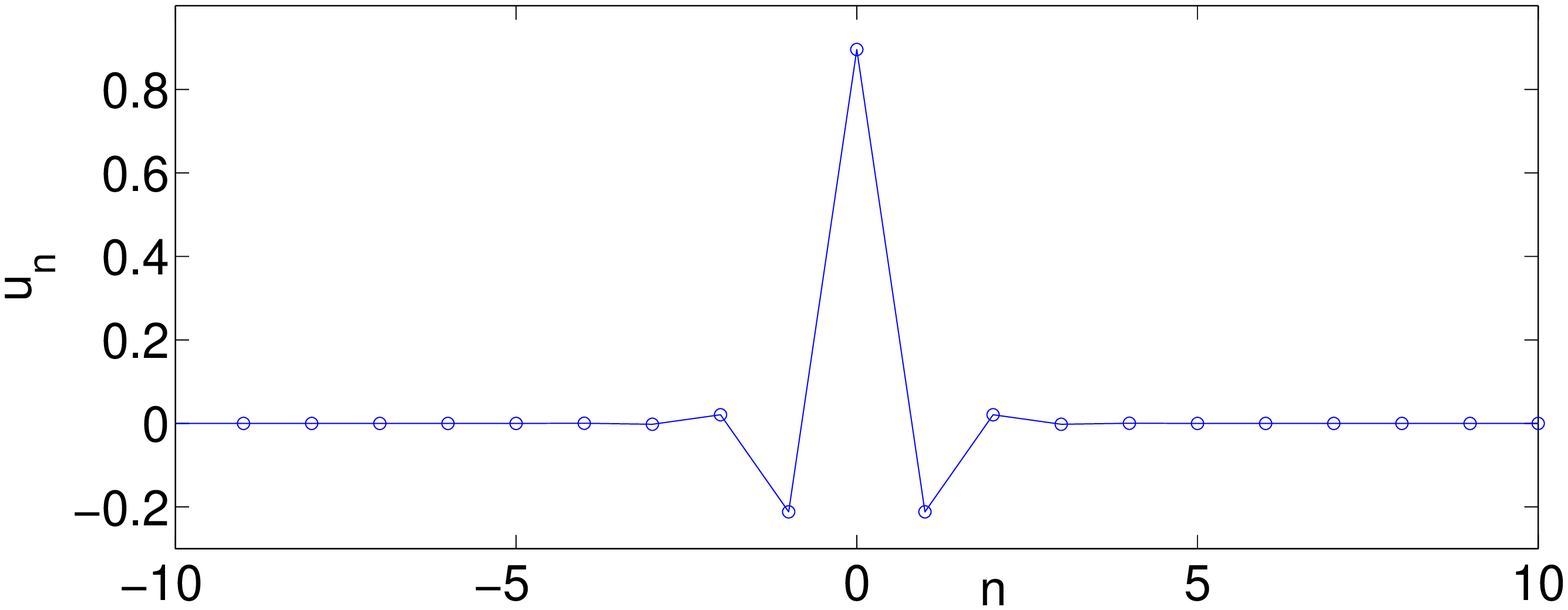}
\end{center}
\begin{center}
\includegraphics[width=0.45\textwidth]{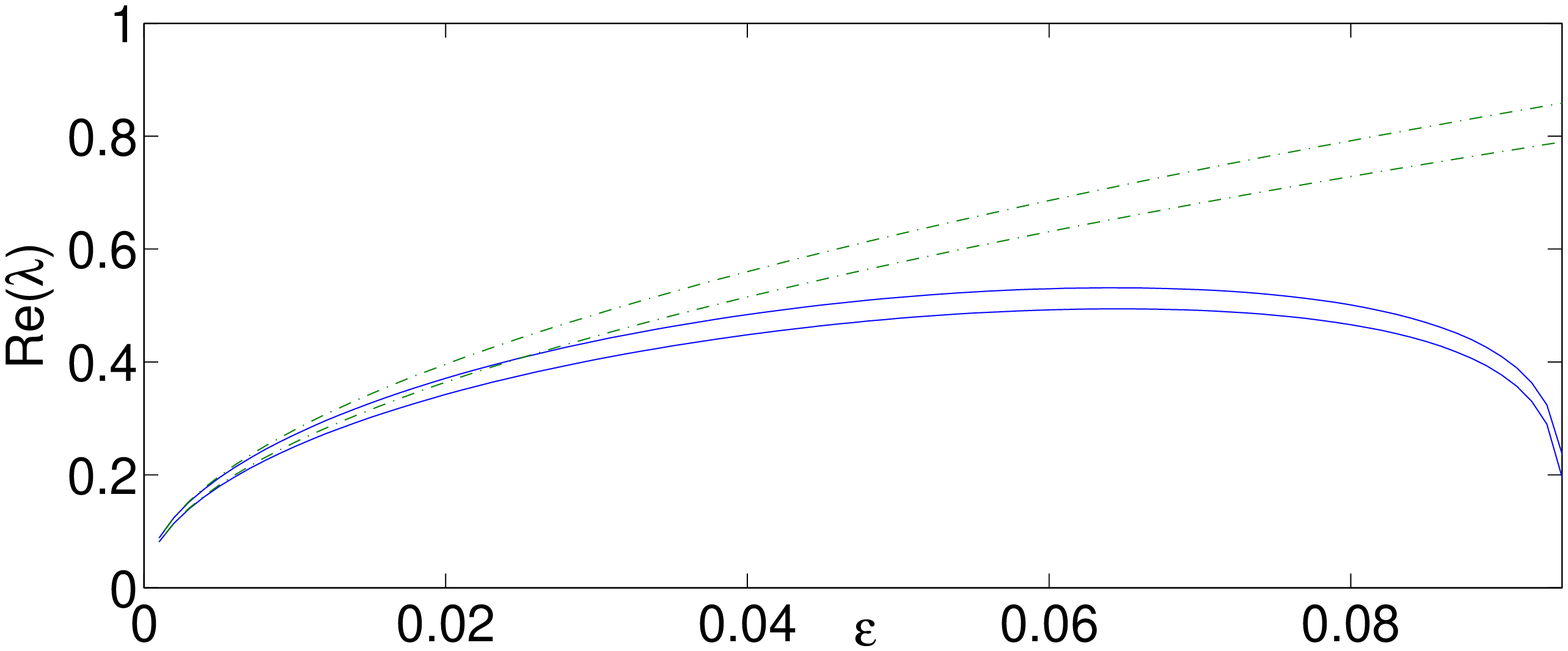}
\includegraphics[width=0.45\textwidth]{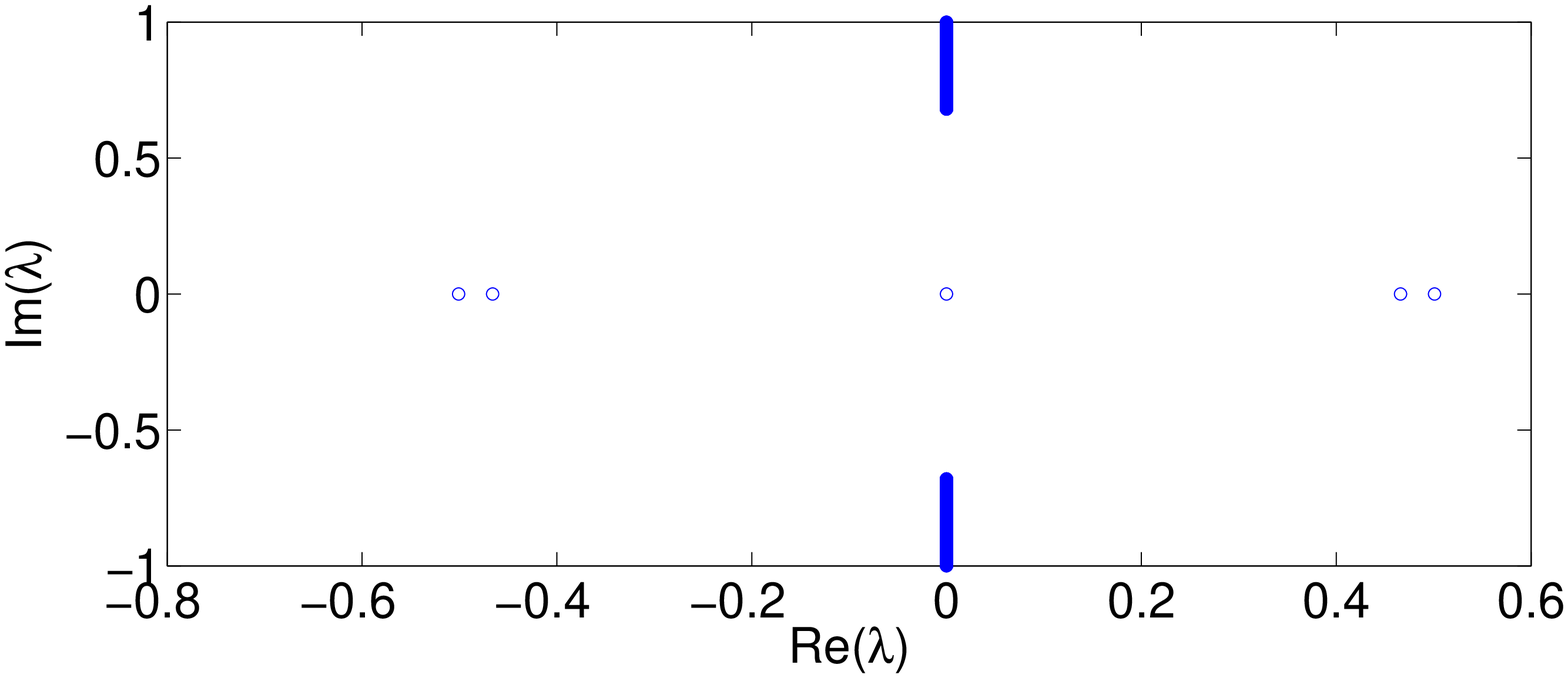}
\end{center}
\caption{Same as for all the previous branches, but now for the 
configuration of the form $(0,\dots,0,-\sqrt{\frac{1}{g(-1)}},\sqrt{\frac{1}{g(0)}},-\sqrt{\frac{1}{g(1)}},0,\dots0)$ at the AC limit, which bears
two real eigenvalue pairs and terminates at $\epsilon=0.095$.}
\label{kps_fig5}
\end{figure}

\begin{figure}[!ht]
\begin{center}
\includegraphics[width=0.45\textwidth]{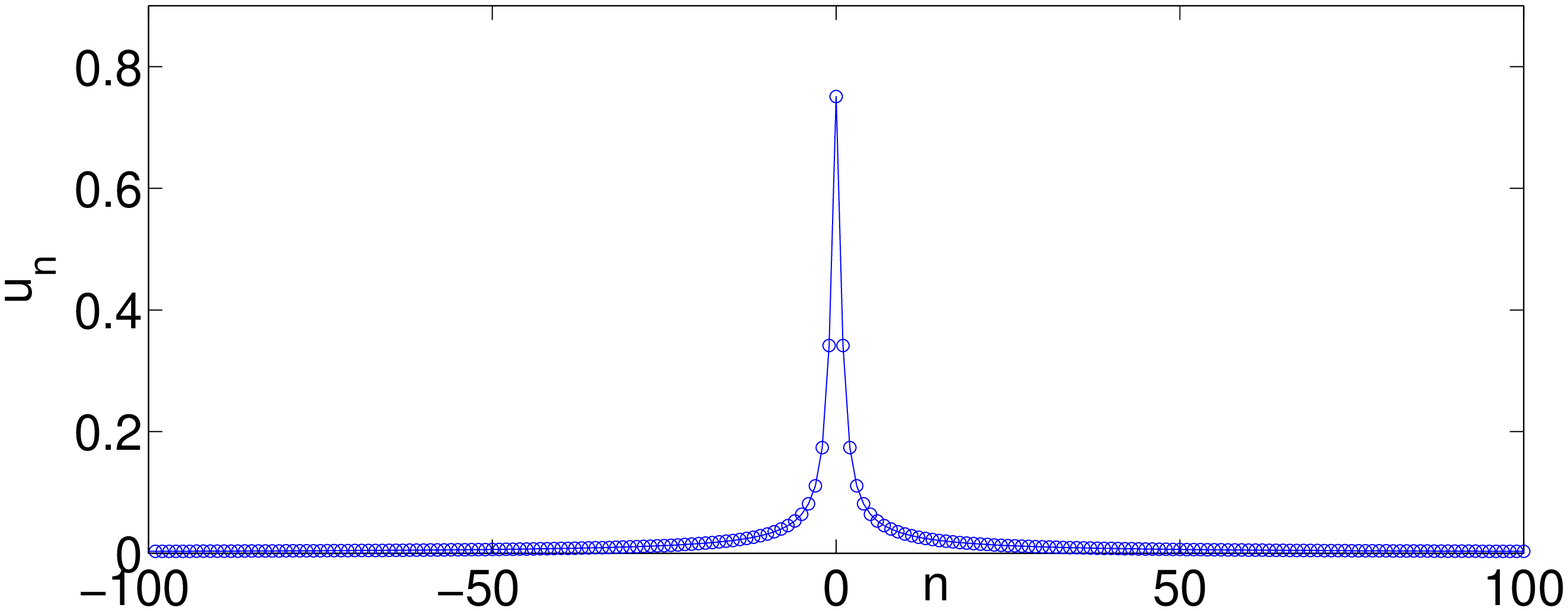}
\includegraphics[width=0.45\textwidth]{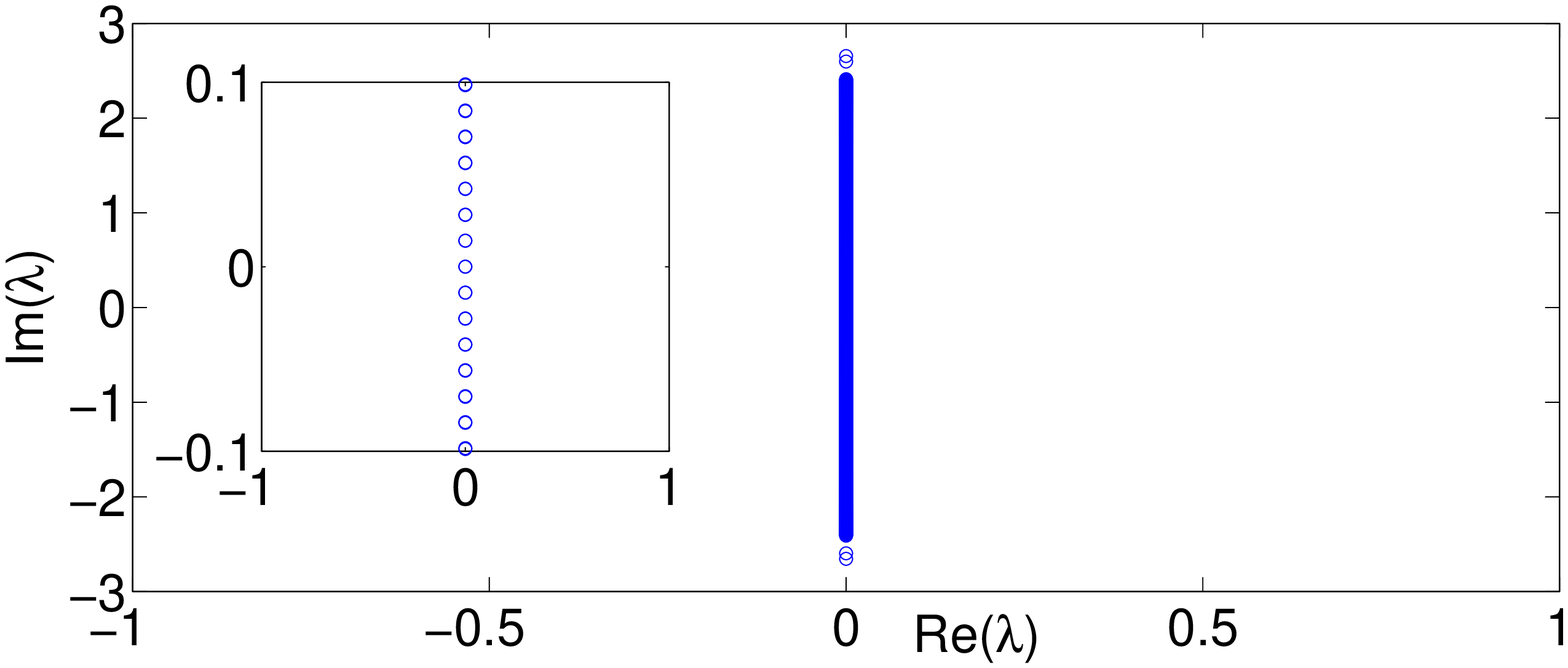}
\end{center}
\caption{The profile (left panel) and the stability (right panel)
again for $\epsilon=0.08$ but now for the ``extended'' configuration
in which all sites are excited at $\epsilon=0$ according to:
$v_n= \sqrt{1/g(n)}$. The configuration is generically stable
as is also illustrated by the spectrum and its zoom-in inset
in the right panel.}
\label{kps_fig6}
\end{figure}

We now briefly explore the ``extended'' branch in which at
$\epsilon=0$, all the sites are excited. Given the
algebraically growing structure of the nonlinear prefactor,
the form $v_n= \sqrt{1/g(n)}$ provides a decaying wave profile.
Firstly, it is interesting to note here that for all the values
of $\epsilon$ considered in our computation, this profile was
found to persist, suggesting, similarly to~\cite{malom10}, that
this solution may persist all the way to the continuum limit.
It is generally worthwhile to iterate here that we found that
configurations with progressively larger support were found
to persist for larger intervals of $\epsilon$ values. This
is entirely contrary to what is known e.g. for the standard
homogeneous focusing case (see~\cite{konoalfi} for a relevant 
discussion), where the more localized configurations  are
the ones eventually persisting all the way to the limit,
while all others disappear through suitable bifurcations.
For the relevant extended solution presented in Fig.~\ref{kps_fig6},
it is worthwhile to also touch upon its spectrum. Interestingly,
since {\it all} the sites are excited at $\epsilon=0$, each
of them is also associated with a zero pair. Hence, all eigenvalues
are initially at the origin and bifurcate from there.
The result is the apparent discrete spectrum in the right panel
of Fig.~\ref{kps_fig6}, whereby the eigenvalue pairs
parabolically grow as $\epsilon$ increases. The detailed stability
properties of such a configuration merit separate investigation,
but suffice it to mention for present purposes that the configuration
was found to be stable for all the considered values of the 
coupling strength.

\begin{figure}[!ht]
\begin{center}
\includegraphics[width=0.45\textwidth]{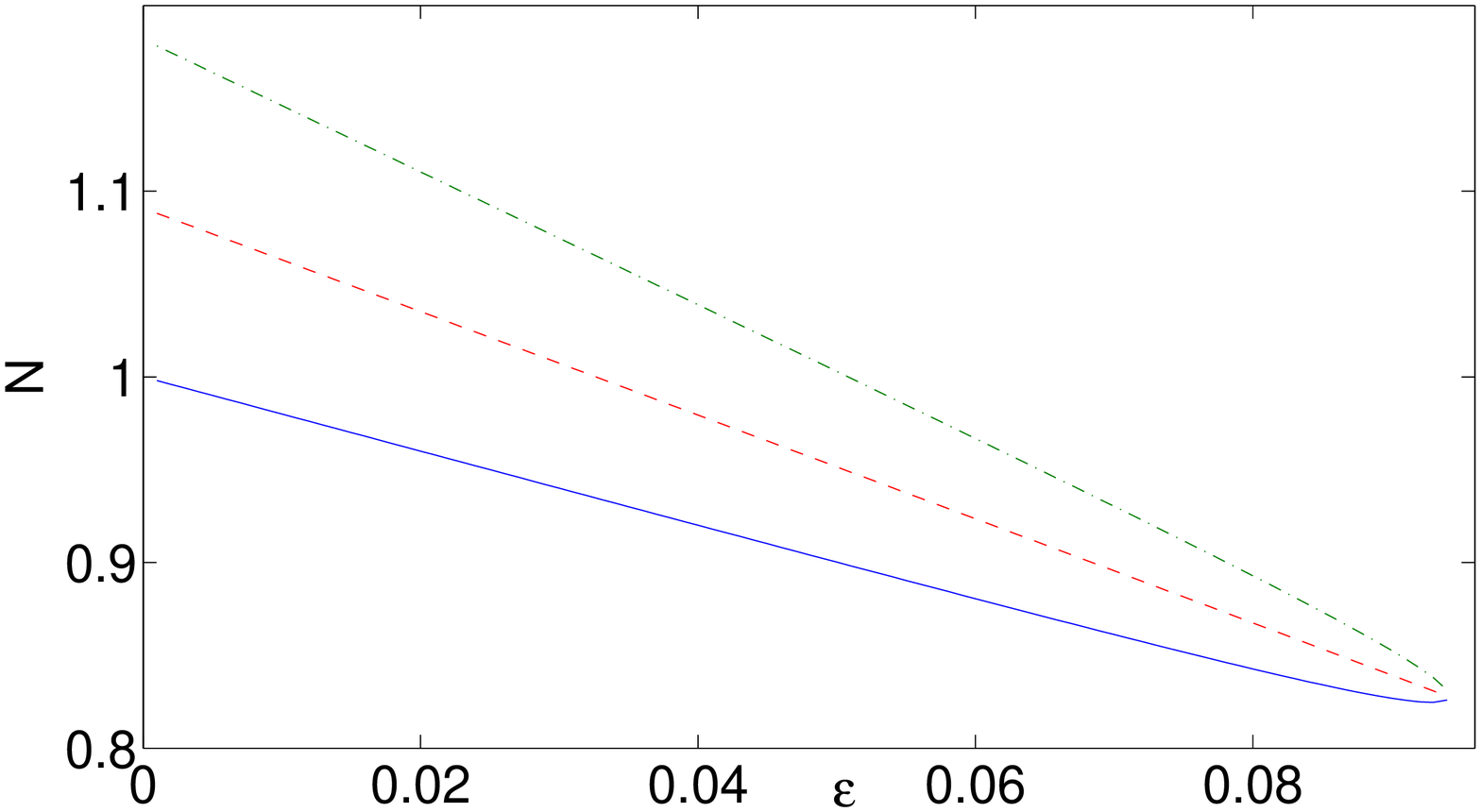}
\includegraphics[width=0.45\textwidth]{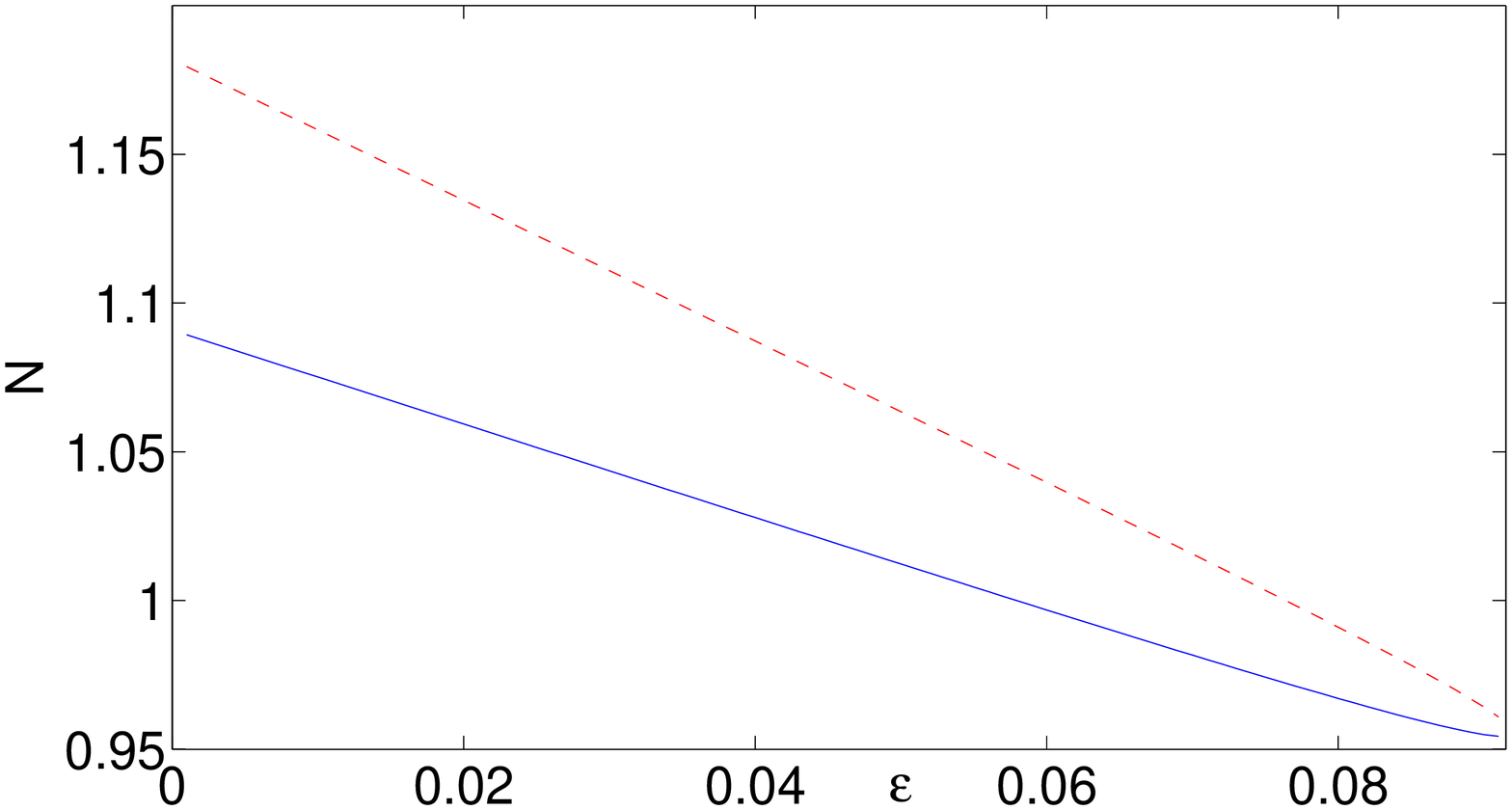}
\end{center}
\caption{The left panel shows the collision of the single-site branch
$(0,\dots,0,\sqrt{\frac{1}{g(0)}},0,\dots0)$ (blue solid line) with
the two out-of-phase site branch $(0,\dots,0,\sqrt{\frac{1}{g(0)}},-\sqrt{\frac{1}{g(1)}},0,\dots0)$ (red dashed line) 
and the three out-of-phase site branch
 $(0,\dots,0,-\sqrt{\frac{1}{g(-1)}},\sqrt{\frac{1}{g(0)}},-\sqrt{\frac{1}{g(1)}},0,\dots0)$ (green dash-dotted line), through monitoring the dependence
of their respective powers ${\cal N}=\sum_n |u_n|^2$ as a function
of $\epsilon$. The right panel is similar but now for the collision 
of  $(0,\dots,0,\sqrt{\frac{1}{g(0)}},\sqrt{\frac{1}{g(1)}},0,\dots0)$
(blue solid line)
with $(0,\dots,0,-\sqrt{\frac{1}{g(-1)}},\sqrt{\frac{1}{g(0)}},\sqrt{\frac{1}{g(1)}},0,\dots0)$ (red dashed line).}
\label{kps_fig6_update}
\end{figure}

A summary of the different types of states that are examined above
is provided in Table~\ref{t1}, together with the bifurcations
leading to their termination and the associated (approximate) critical
points. Additionally, in Fig.~\ref{kps_fig6_update}, we offer
an alternative diagnostic that can also be meaningfully used
to detect the relevant bifurcations and branch collisions.
In particular, we show ${\cal N}= \sum_n |u_n|^2$ as a function 
of $\epsilon$, which allows to monitor the continuation of
the solutions for different values of coupling
strength parameter. The left panel shows the branches
A, C, and E of the table, namely the single-site, 
two out-of-phase and three adjacent out-of-phase sites
which collide in the double pitchfork bifurcation around
$\epsilon=0.095$, while the right panel illustrates branches
B and E, namely the in-phase two-site branch and the one
with the form $(0,\dots,0,-\sqrt{\frac{1}{g(-1)}},\sqrt{\frac{1}{g(0)}},\sqrt{\frac{1}{g(1)}},0,\dots0)$ near the $\epsilon=0$ limit, which, in turn,
collide and disappear in a saddle-center bifurcation around
$\epsilon=0.091$.

\begin{figure}[!ht]
\begin{center}
\includegraphics[width=0.45\textwidth]{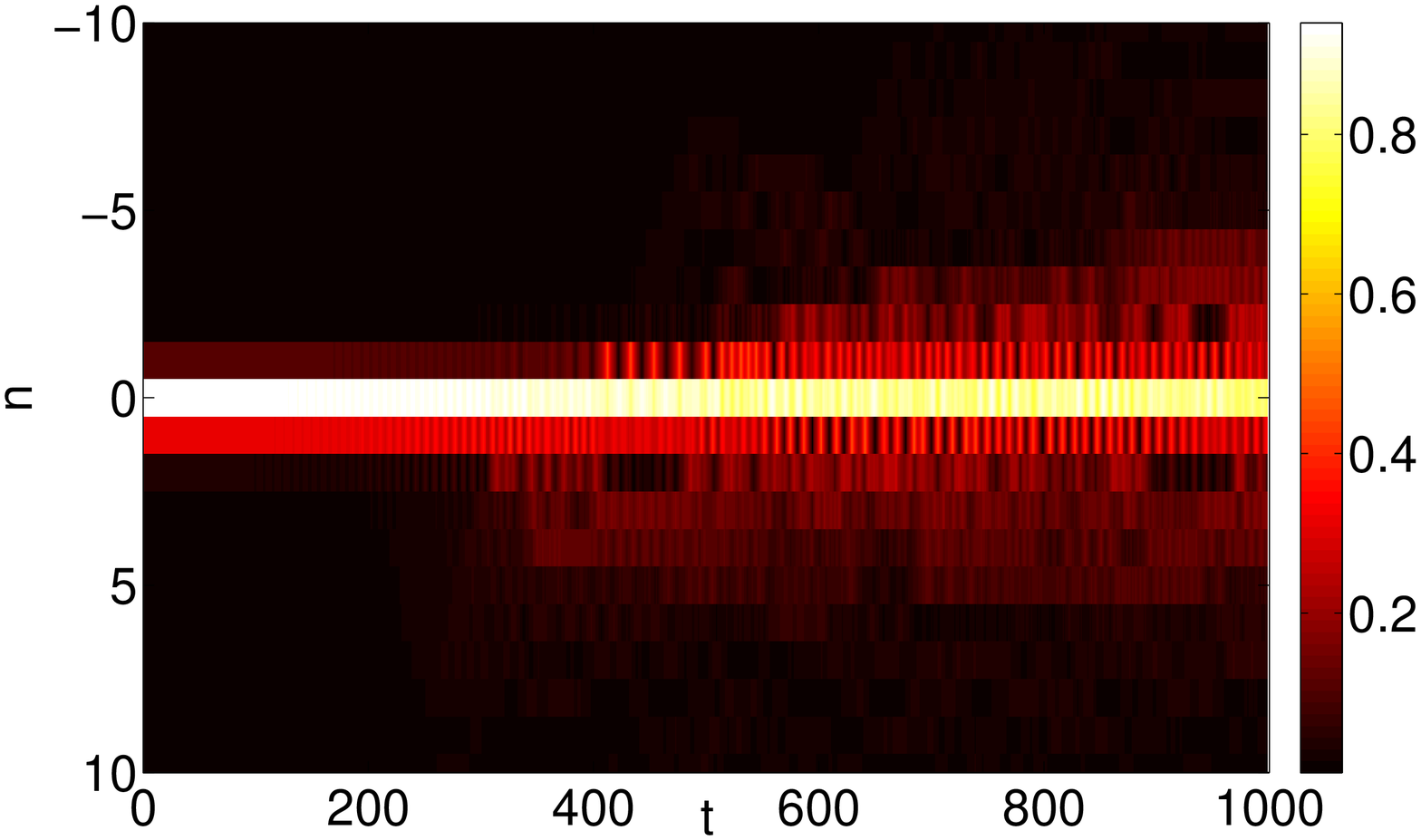}
\includegraphics[width=0.45\textwidth]{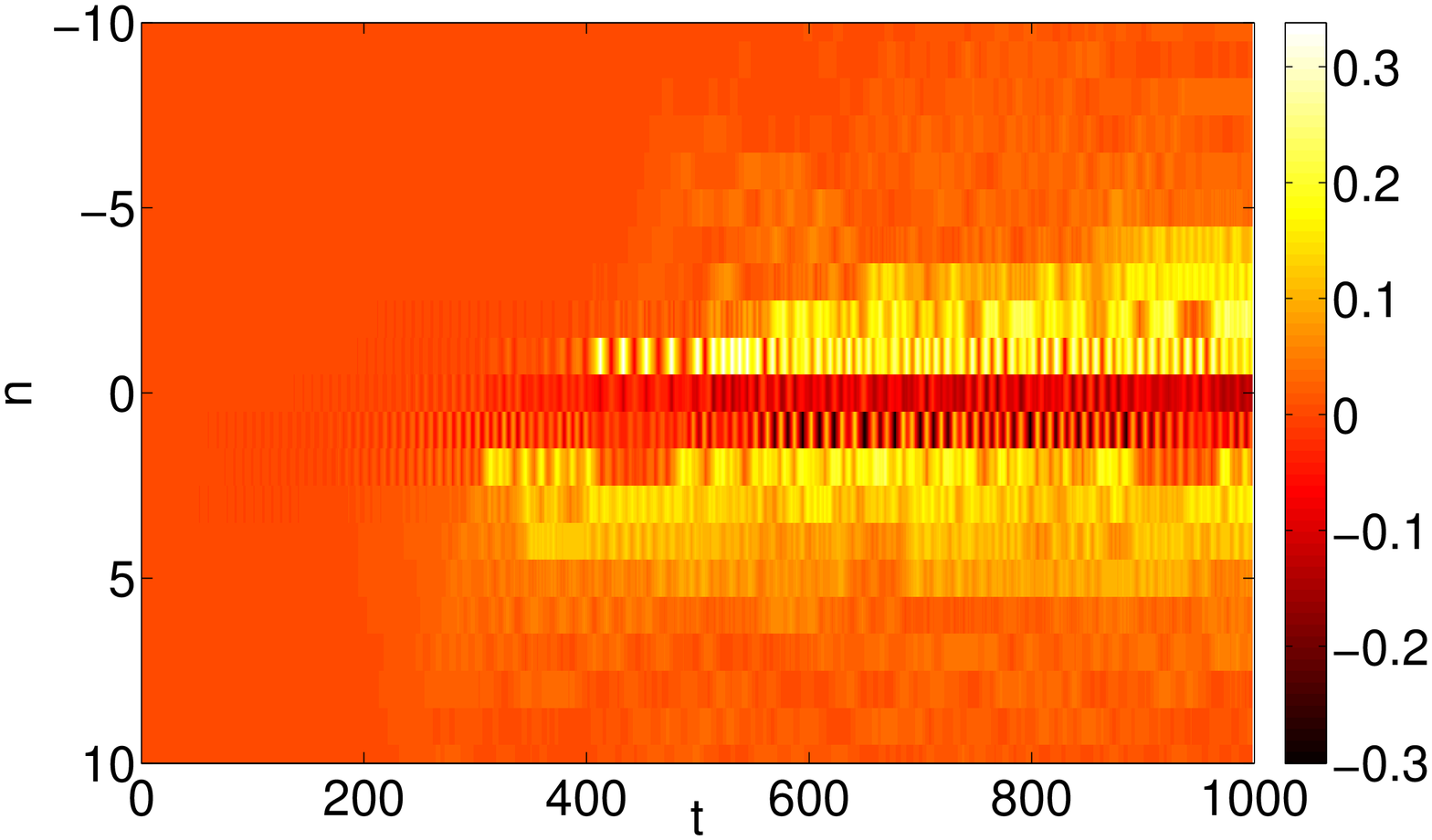}
\end{center}
\caption{The left panel shows the space ($n$)- time ($t$)
evolution of the modulus of the two-site, in-phase
solution (of the form $(0,\dots,0,\sqrt{\frac{1}{g(0)}},\sqrt{\frac{1}{g(1)}},0,\dots0)$ at the AC limit). The right panel shows the difference of
the magnitude of the solution from the magnitude of its corresponding
initialization. The coupling strength here is chosen as
$\epsilon=0.08$.}
\label{kps_fig7}
\end{figure}

\begin{figure}[!ht]
\begin{center}
\includegraphics[width=0.45\textwidth]{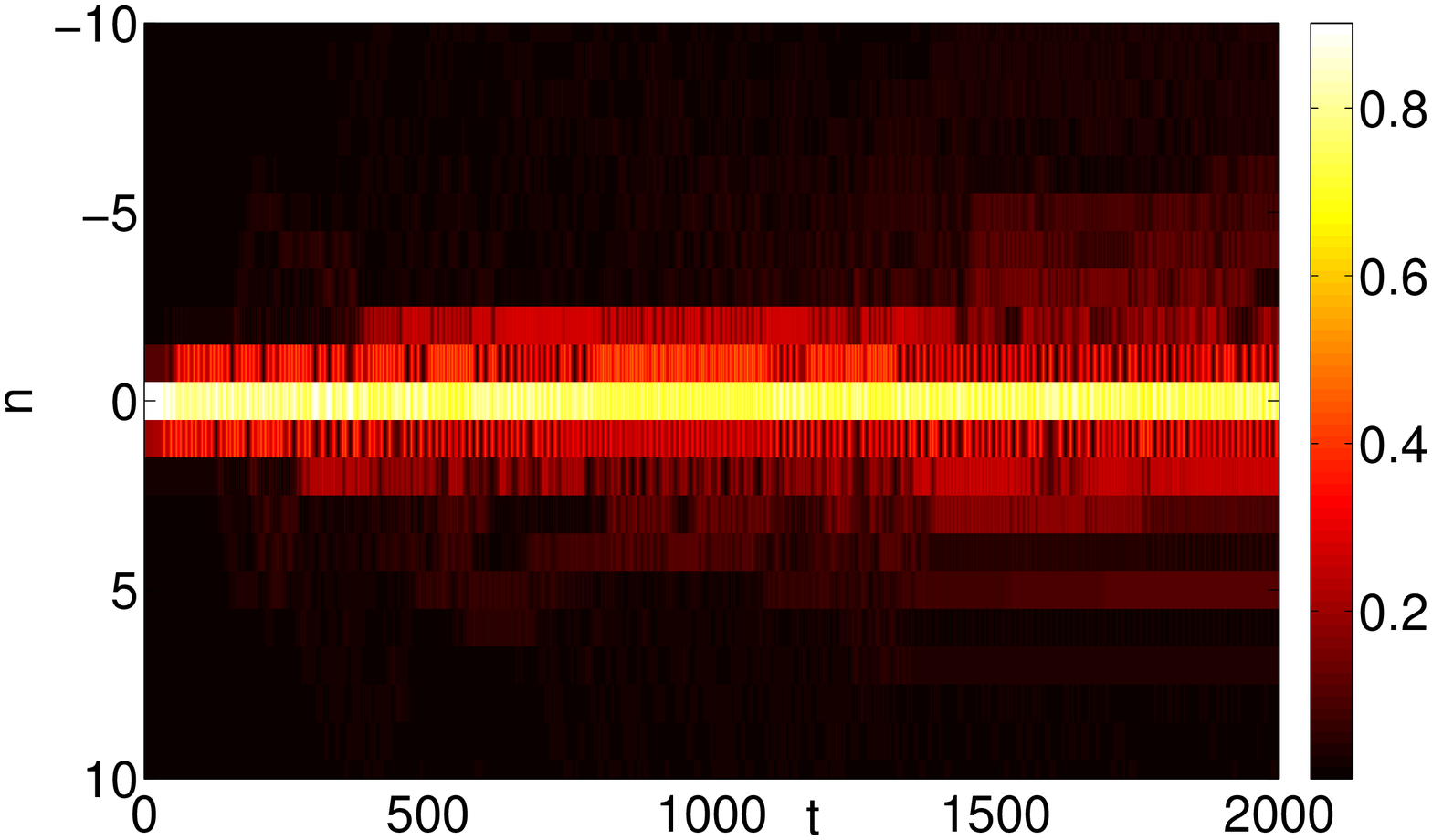}
\includegraphics[width=0.45\textwidth]{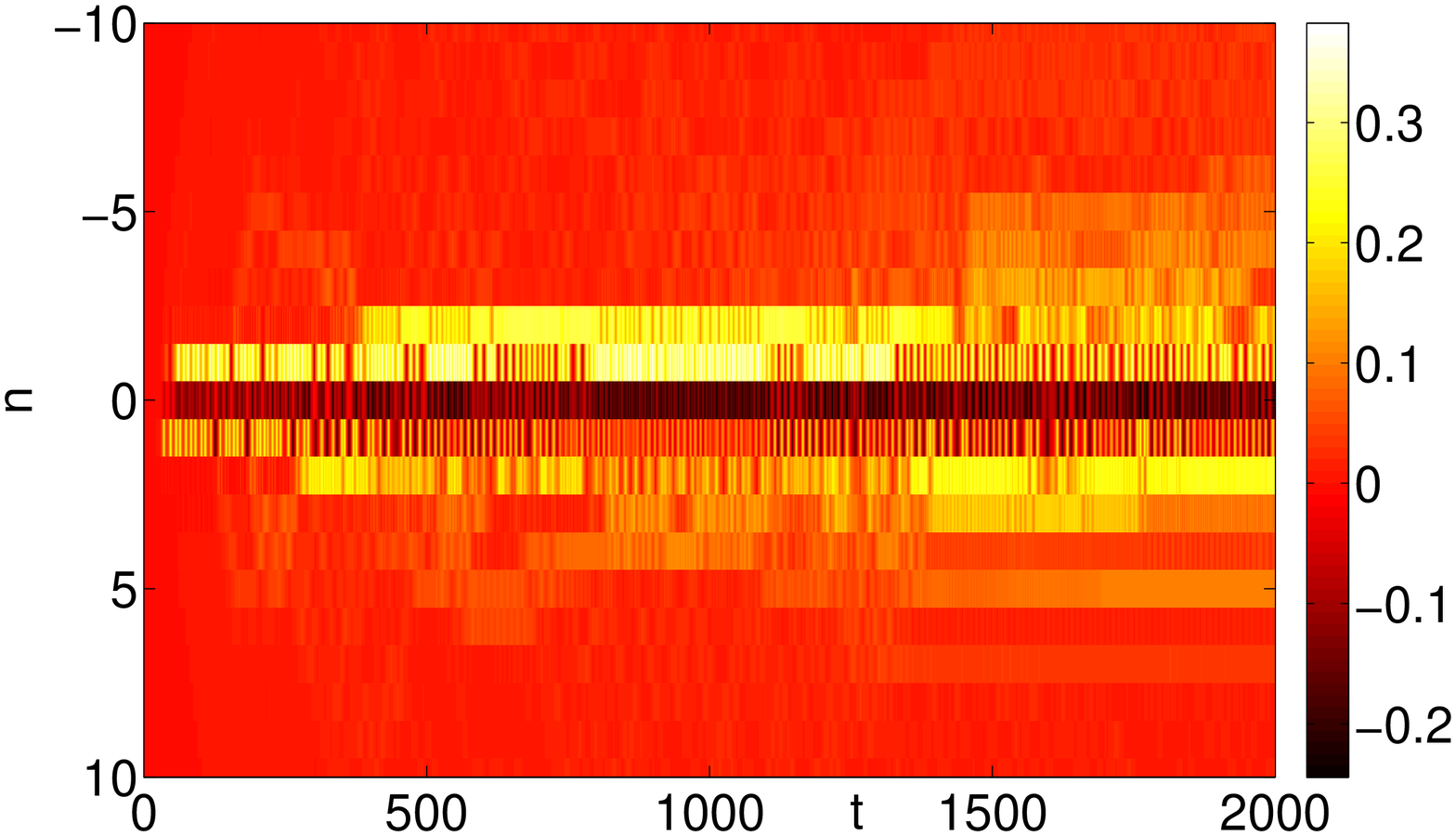}
\end{center}
\caption{Same as in Fig.~\ref{kps_fig7}, but now for the
two-site, out-of-phase solution (of the form
 $(0,\dots,0,\sqrt{\frac{1}{g(0)}},-\sqrt{\frac{1}{g(1)}},0,\dots0)$ at the AC limit).}
\label{kps_fig8}
\end{figure}

\begin{figure}[!ht]
\begin{center}
\includegraphics[width=0.45\textwidth]{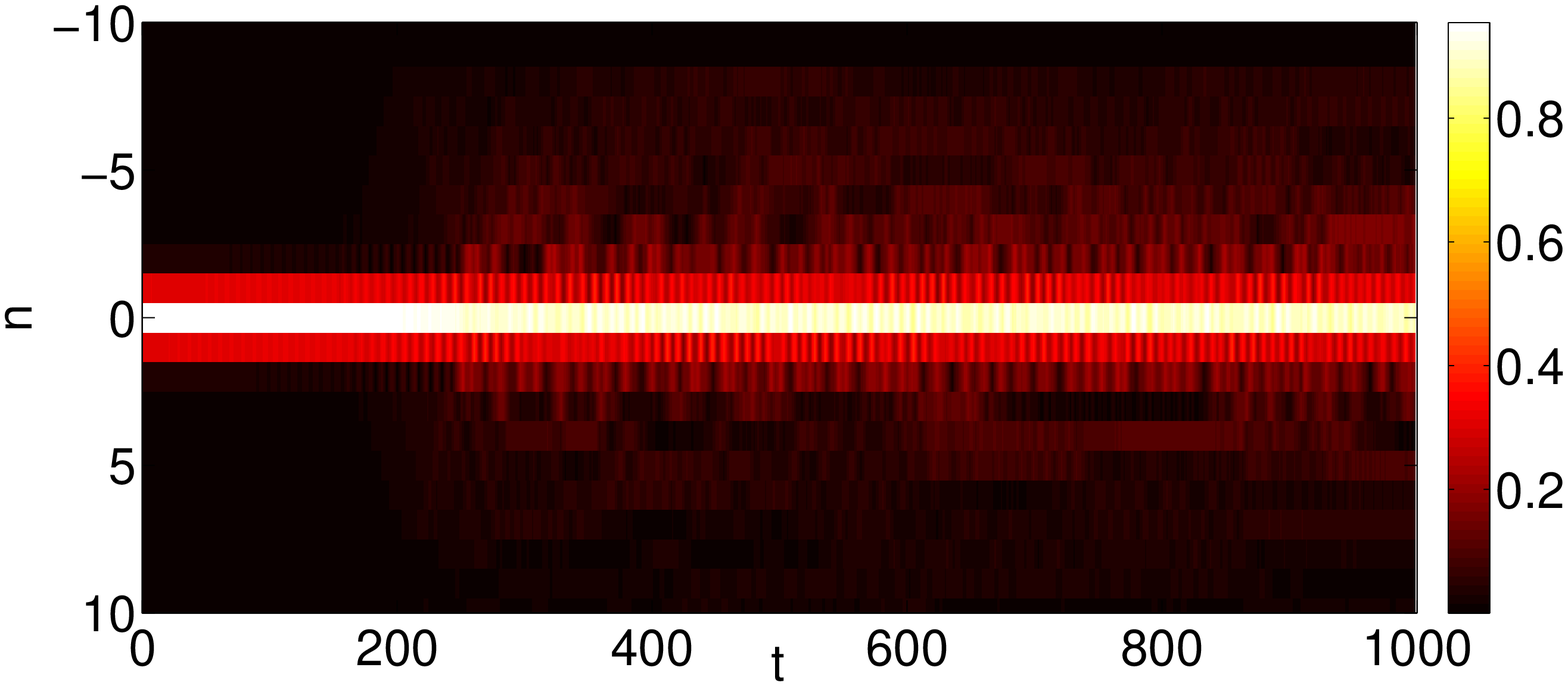}
\includegraphics[width=0.45\textwidth]{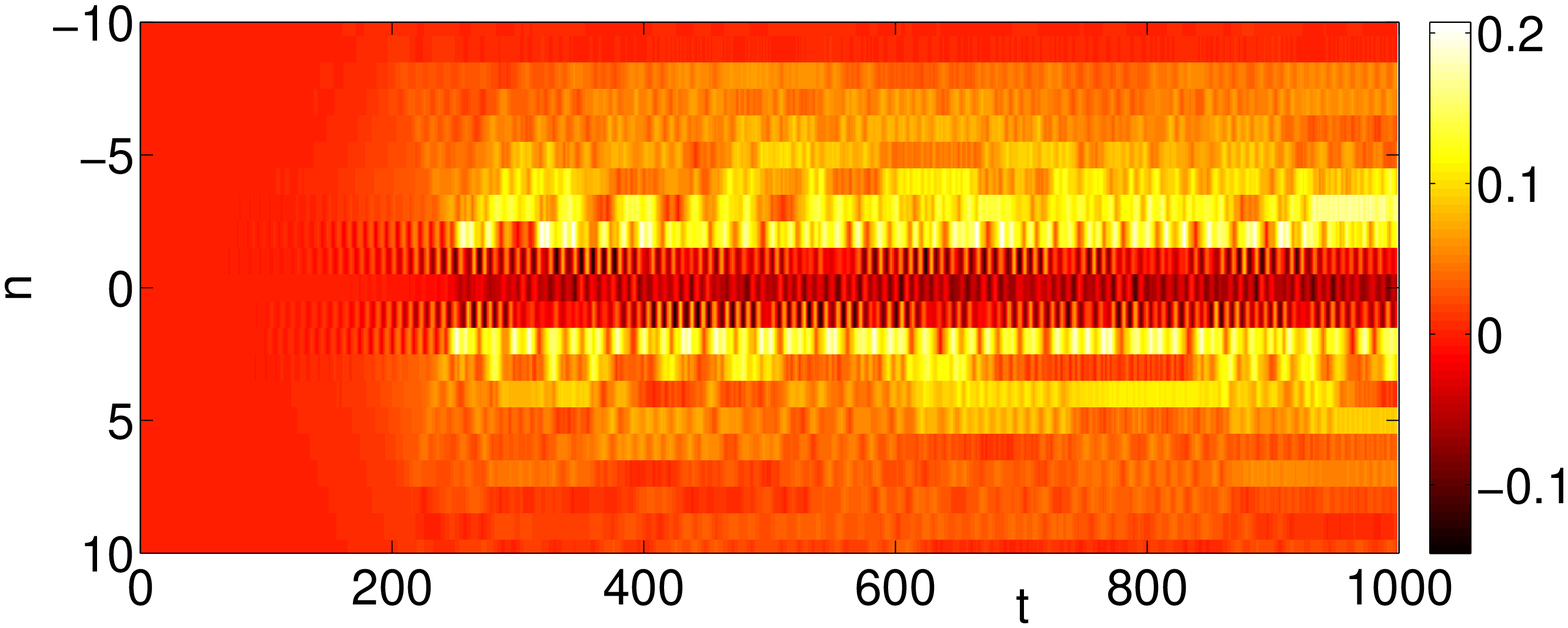}
\end{center}
\caption{Same as the previous two examples but now for the in-phase
three-site solution (of the form $(0,\dots,0,\sqrt{\frac{1}{g(-1)}},\sqrt{\frac{1}{g(0)}},\sqrt{\frac{1}{g(1)}},0,\dots0)$ in the AC limit).} 
\label{kps_fig9}
\end{figure}

Finally, we now turn to the dynamical exploration of the
evolution of the unstable configurations in the space-time
numerical experiments of Figs.~\ref{kps_fig7}, 
\ref{kps_fig8} and~\ref{kps_fig9}. These are all performed
for the case of $\epsilon=0.08$ used previously to showcase
the solution profiles. Given the similarity of the profiles
of the different branches (and the bifurcations elucidated
above), we only show three out of the five few-site excited
branches (recall that the single-site excited branch, as
well as the extended profile branch are stable throughout
their respective regimes of existence). Fig.~\ref{kps_fig7}
illustrates the case of the two-site in-phase excitation
branch, Fig.~\ref{kps_fig8} corresponds to the out-of-phase
two-site excitation, while Fig.~\ref{kps_fig9} is associated
with the three-site in-phase excitation. Recall that the
mixed phase three-site excitation is rather similar in 
profile to the two-site in-phase, as is the three-site out-of-phase
to the two-site out-of-phase for this value of $\epsilon$.

In all three cases, the two panels, respectively, demonstrate
the space-time evolution of the contour of the solution 
magnitude and its difference (again in magnitude) from its
initial spatial profile. The former provides a sense of the
dynamics, while the latter also gives a glimpse of the type
of instability that results in it. Interestingly, in all the
cases we see a rather similar evolution, i.e., over time
while the dynamics does not appear to definitively settle to an 
asymptotic state, it does seem to expand its spatial extent,
lending further support to the idea that configurations with
more excited sites are favored in the present setting. 
On the other hand, we do also detect some differences
between the different cases. In particular, the oscillatory
instabilities of Fig.~\ref{kps_fig7} and Fig.~\ref{kps_fig9},
bear a much weaker growth rate (as is typically the case
for oscillatory instabilities in comparison to exponential
ones), and thus require a far longer (by
an order of magnitude, which roughly mirrors the corresponding
difference in growth rates) time interval to manifest themselves
in comparison to the rapidly developing exponential growth
of  Fig.~\ref{kps_fig8}. In the former cases, the right panel 
appears to also mirror the oscillatory nature of the instability
at its dynamical onset i.e., there is an interval
of oscillatory growth as is expected by the complex nature
of the unstable eigenvalues associated with these
cases. It is also relevant to point out that
in these cases, to seed the instability a random (uniformly
distributed) noise has been added to the initial condition,
while in the case of Fig.~\ref{kps_fig8} this was not necessary
(i.e., numerical round-off error was rapidly --exponentially-- 
amplified in the
latter setting).

\section{Conclusions \& Future Challenges}

In the present work, we have explored a setup of increasing interest
in the theory of nonlinear waves in lattices and continua, namely
the emergence of bright solitary waves in defocusing nonlinear
media, in the presence of a spatially inhomogeneous nonlinearity
profile. Our specific interest here was to explore the lattice
setting, of particular potential relevance to waveguide applications.
The perspective utilized was that of the so-called anti-continuous limit
which enabled a systematic theoretical analysis, perturbatively close
to that limit. This provided not only a roadmap on the available
coherent structures, but more importantly a handle on their expected
stability properties. The analytical results obtained by means of
this approach were fully corroborated by detailed numerical existence
and spectral computations. The latter additionally revealed the
bifurcation type scenarios that emerge, as well as provided an
understanding on which states may be favored in such a setting.
We also used a number of proof-of-principle numerical simulations
in order to explore the dynamical evolution of potentially unstable
states.

We believe that these efforts will provide further insight
on the relevant phenomenology and will also give a significant amount
of motivation for their exploration in experimental setups in 
nonlinear optics that presently appear to be well within 
reach. Further theoretical efforts could focus on a variety of
settings. It would be interesting for example to provide
an analytical characterization of the spectral operators and the
stability of the extended (stable) state that we discussed herein,
as well as to explore the similarities and differences
(existence, stability and dynamics-wise) of different 
``profiles'' of the inhomogeneous nonlinearity, such as the
exponential one previously studied in~\cite{malom10} vs. 
power-law (as e.g. in the case example considered herein).
Our analysis, to the extent possible herein, was kept very general,
and clearly some features (like the decay of the extended state)
will accordingly differ, but if some broad qualitative statements
could be made along these lines, it would be especially useful
(including in designing relevant experiments).
Lastly, and perhaps most importantly exploring such systems in 
higher dimensions and identifying the impact on such inhomogeneous
nonlinearities on different kinds of structures, including vortical
ones would be an especially relevant theme for future investigations.
Efforts in this direction have been recently initiated in 2d settings (see 
e.g.~\cite{inhdef_2d}) and may well be relevant to extend also to
3d case examples.

\vspace{5mm}

{\it Acknowledgements.} 
We gratefully acknowledge the support of NSF-DMS-0806762
and NSF-DMS-1312856, NSF-CMMI-1000337, as well as from
the AFOSR under grant FA9550-12-1-0332, the
Binational Science Foundation under grant 2010239, from the
Alexander von Humboldt Foundation and the ERC under FP7, Marie
Curie Actions, People, International Research Staff
Exchange Scheme (IRSES-605096).
P.G.K.’s work at Los Alamos is supported in part by the U.S. Department
of Energy. D.K. acknowledges support from DFG Ki482/16-1.

\end{document}